\documentclass[12pt]{article}
\usepackage[paper=letterpaper,margin=1in]{geometry}
\usepackage{amsmath,amssymb,bm,mathtools}
\numberwithin{equation}{section}
\numberwithin{equation}{subsection}
\renewcommand{\theequation}{\ifnum\value{subsection}>0\thesubsection\else\thesection\fi.\arabic{equation}}
\usepackage{xcolor}
\usepackage[normalem]{ulem}
\usepackage[numbers,sort&compress]{natbib}
\usepackage{cancel}
\usepackage{tikz}
\usetikzlibrary{decorations.pathmorphing, arrows.meta}
\usepackage{booktabs,array}
\usepackage[colorlinks=true,linkcolor=blue,citecolor=blue,urlcolor=blue]{hyperref}
\newcommand{\be}{\begin{equation}}
\newcommand{\ee}{\end{equation}}
\newcommand{\bea}{\begin{eqnarray}}
\newcommand{\eea}{\end{eqnarray}}

\newcommand{\bra}[1]{\left<{#1}\right|}
\newcommand{\ket}[1]{\left|{#1}\right>}

\newcommand{\dd}{\mathrm{d}}
\newcommand{\ii}{\mathrm{i}}
\newcommand{\cH}{\mathcal{H}}
\newcommand{\cS}{\mathcal{S}}
\newcommand{\cI}{\mathcal{I}}
\newcommand{\rh}{r_{\mathrm h}}

\newcommand{\eps}{\epsilon}
\newcommand{\vev}[1]{\left\langle #1 \right\rangle}

\newcommand{\daggerfootnote}[1]{%
    \renewcommand{\thefootnote}{\fnsymbol{footnote}}%
    \footnote[2]{#1}
    \renewcommand{\thefootnote}{\arabic{footnote}}%
}

\makeatletter
\@ifundefined{ruledtabular}{%
  \newenvironment{ruledtabular}{}{}%
}{}
\makeatother
\providecommand{\colrule}{\hline}

\tikzset{
  causal horizon/.style={thick,dashed},
  causal singularity/.style={thick,decorate,decoration={snake,amplitude=2pt,segment length=6pt}},
  causal boundary/.style={thick},
  causal context/.style={opacity=0.3},
  causal region/.style={fill=black!5,draw=none},
  causal exterior/.style={fill=blue!10,draw=none}
}

\makeatletter
\def\@hangfrom@appendix#1#2#3{#1\@if@empty{#2}{#3}{#2\@if@empty{#3}{}{\quad #3}}}

\begin{document}

\thispagestyle{empty}

\vspace*{.5cm}
\begin{center}

{\bf {\LARGE Near-horizon soft theorems\\[8pt] from local Ward identities }\\
\vspace{1cm}}

\begin{center}

{\bf Jin-Peng Zhuge$^{a}$ and Peng Cheng$^{a}$\daggerfootnote{p.cheng.nl@outlook.com}}\\
\bigskip \rm
  
\bigskip a) Center for Joint Quantum Studies and Department of Physics, \\School of Science, Tianjin University, Tianjin 300350, China\\

\rm
  \end{center}

\vspace{1.5cm} {\bf Abstract}
\end{center}
\begin{quotation}
We develop a unified formulation of leading soft photon theorems for nondegenerate static spherical horizons. By fixing the photon mode normalization from the horizon symplectic form, we obtain a common angular soft factor whose sign follows from the boundary orientation. We determine the geometric coefficients in the local angular and momentum representations for a chosen radial normalization. The charge balance includes horizon endpoint data and contributions from other boundaries. This formulation brings black hole and cosmological horizons into a common description. Applying the same construction to Yang-Mills fields yields the corresponding tree-level soft gluon theorem. These results identify the common soft structure underlying different horizon geometries.
\end{quotation}

\vspace{1cm}

\setcounter{page}{0} \setcounter{tocdepth}{2} \setcounter{footnote}{0}


\newpage \noindent\rule[-10pt]{\textwidth}{0.05em}\par \tableofcontents \noindent\rule[-10pt]{\textwidth}{0.05em}\par

\section{Introduction}
\label{sec-introduction}

The low-energy behavior of gauge theory was constrained long before the modern language of asymptotic symmetry was developed. Low showed that the first terms in soft-photon emission are fixed by the nonradiative process, and Weinberg established the universal leading photon and graviton factors for relativistic scattering \cite{Low:1958sn,Weinberg:1965nx}. These results identify a sector of the amplitude that is insensitive to short-distance dynamics but remains sensitive to the charges and momenta of the hard particles. The same leading structure appears for Yang-Mills theory, with color generators replacing Abelian charges \cite{Strominger:2013lka}. The modern interpretation relates soft factors to Ward identities associated with asymptotic symmetries \cite{Strominger:2013jfa,He:2014laa,He:2014cra,Campiglia:2015qka,Kapec:2015ena,Strominger:2017zoo}. Electromagnetic memory supplies a counterpart of the same zero-frequency sector \cite{Bieri:2013hqa,Pasterski:2015zua}. This relation does not remove the infrared subtleties of scattering with charged particles. Physical charged states carry long-range fields, and formulations based on dressed asymptotic states or memory sectors clarify why a naive Fock-space amplitude is not the only useful object \cite{Kulish:1970ut,Gabai:2016kuf,Prabhu:2024zaf}.

Extending this picture from infinity to a null boundary requires additional care. Surface charges on radiating boundaries obey flux laws rather than ordinary time-independent conservation laws, and their Hamiltonian interpretation depends on boundary and corner data \cite{Iyer:1994ys,Wald:1999wa,Barnich:2001jy,Donnelly:2016auv,Chandrasekaran:2018aop,Harlow:2019yfa,Ball:2024yze}. For a horizon, a local constraint can determine a soft insertion, but it cannot by itself specify how the two horizon branches are glued or what happens to charge on the rest of the spacetime boundary. Those issues are part of the definition of the Ward identity rather than optional global decorations. Black hole horizons have consequently become a natural setting in which to test the infrared symmetry picture. Large horizon symmetries and their charges motivated the proposal of soft black hole hair and a broader analysis of nonextremal near-horizon symmetry algebras \cite{Hawking:2016msc,Donnay:2015abr,Donnay:2016ejv}. Low-frequency absorption can also be organized as a large-gauge conservation law \cite{Avery:2016mrt}. At the same time, dressed-state analyses show that the existence of soft charges alone may not establish a new channel for storing hard information \cite{Mirbabayi:2016axw,Bousso:2017dny}. One therefore should use the horizon Ward identity as a precise statement about boundary gauge data.

Earlier works derived soft photon and soft gluon relations near the Schwarzschild horizon \cite{Cheng:2022xyr, Cheng:2022xgm}, while related studies investigated the de Sitter (dS) soft photon theorem and linearized gravity near a horizon \cite{Mao:2023rca,Mao:2023dsy}. These results motivate a unified description of soft theorems for different types of horizons. In particular, it remains to establish how black hole and cosmological horizons fit into a common framework, which features of their soft relations are universal, and how differences in horizon geometry and boundary orientation enter their formulation. In this work, we develop such a framework for probe gauge fields near nondegenerate static spherical horizons. We consider a relatively general background throughout the neighbourhood used in the characteristic analysis. Including the electric endpoint data in the horizon phase space allows us to construct the large-gauge generator and derive the corresponding Ward identity. With the soft mode normalized by the horizon symplectic form, the leading soft photon theorem takes a common angular form for black hole and cosmological horizons, with the relative sign fixed by boundary orientation. We then express the theorem in angular and local momentum variables within the same isotropic mode convention. Matching the horizon states to local massless momentum modes relates the two soft factors by a momentum contraction, with a geometric coefficient fixed by the horizon data.

This formulation brings the Schwarzschild and dS results into a common description and applies to more general backgrounds, as illustrated by Reissner-Nordstr\"om (RN) and Schwarzschild-de Sitter (S-dS) geometries. The same construction yields a tree-level extension to soft gluons for a probe Yang-Mills field. The Ward identity also makes explicit the charge exchanged with other boundary components, thereby specifying when the horizon soft relation holds. Under the stated matching conditions, the homogeneous theorem follows in sectors with no change in that auxiliary charge; otherwise, the charge difference supplies an additional term. The resulting framework identifies the shared soft structure of different horizons while retaining the boundary conditions required for its physical interpretation.

We derive horizon soft Ward identities for probe gauge fields on different types of background geometries and horizons. The remainder of the paper is organized as follows. Section~\ref{sec-horizon-charges} derives the completed charge and the global balance law. Section~\ref{sec-soft-theorem} derives the horizon soft theorem and its angular and local momentum representations. Section~\ref{sec-applications} compares representative static backgrounds and gives the tree-level Yang-Mills extension. Section~\ref{sec-discussion} summarizes the invariant content and the limitations of the local construction. Appendix~\ref{app-normalization} collects the normalization checks used in the main text.

\section{Horizon charges}
\label{sec-horizon-charges}

Deriving the horizon soft photon theorem from large gauge symmetry requires identifying the relevant charges and the conservation law they obey. In this section, we construct these charges from the near-horizon Maxwell equations, retaining the endpoint electric data. Matching the past and future horizon data and accounting for charge exchanged with other boundaries then leads to the Ward identity used below.

\subsection{Near-horizon Maxwell equations}

We first specify the horizon data and derive the Maxwell constraint that relates their evolution to the matter flux. Consider a static, spherically symmetric background
\begin{equation}
\dd s^2=-f(r)\dd t^2+h(r)\dd r^2+2r^2\gamma_{z\bar z}\dd z\dd\bar z, \label{eq-static}
\end{equation}
in stereographic coordinates. Here $r$ is the areal radius and $\gamma_{z\bar z}=2/(1+z\bar z)^2$. The functions $f$ and $h$ are independent and positive in the static region. At the selected nondegenerate horizon $r=\rh$, $f$ has a simple zero and
\begin{equation}
c(r)=\sqrt{f(r)h(r)},\qquad c_h=c(\rh)>0,
\end{equation}
extends smoothly across the horizon. In retarded coordinates $u=t-r_*$, with
\begin{equation}
r_*=\int^r\sqrt{\frac{h(s)}{f(s)}}\dd s,
\end{equation}
the metric takes the form
\begin{equation}
\dd s^2=-f(r)\dd u^2-2c(r)\dd u\dd r+2r^2\gamma_{z\bar z}\dd z\dd\bar z. \label{eq-null-metric}
\end{equation}
The coordinate transformations are collected in Appendix~\ref{sec-coordinates}.

Writing $x=r-\rh$, the near-horizon expansions are
\begin{equation}
f(r)=f'(\rh)x+\mathcal O(x^2),\qquad h(r)=\frac{c_h^2}{f'(\rh)x}+\mathcal O(1).
\end{equation}
On the static side, $x/f'(\rh)>0$, and the proper distance from the horizon is
\begin{equation}
\ell=2c_h\sqrt{\frac{x}{f'(\rh)}}+\mathcal O(|x|^{3/2}).
\end{equation}
The leading static geometry is therefore
\begin{equation}
\dd s^2=-\left[\frac{f'(\rh)^2}{4c_h^2}\ell^2+\mathcal O(\ell^4)\right]\dd t^2+\dd\ell^2+\left[\rh^2+\mathcal O(\ell^2)\right]\dd\Omega_2^2.
\end{equation}
The retarded chart provides a regular description of the corresponding horizon branch, whose spherical sections have finite area $4\pi\rh^2$.

On this fixed background, we consider a probe Maxwell field coupled to charged matter. We keep the coupling in the kinetic term and define the current by variation of the same action
\begin{equation}
S[A,\Phi]=-\frac{1}{4g^2}\int\dd^4x\sqrt{-g}F_{\mu\nu}F^{\mu\nu}+S_{\mathrm m}[A_\mu,\Phi].
\end{equation}
The field equation and current convention are
\begin{equation}
\nabla_\mu F^{\mu\nu}=g^2j^\nu,\qquad j^\nu=-\frac{1}{\sqrt{-g}}\frac{\delta S_{\mathrm m}}{\delta A_\nu}.
\end{equation}
For a scalar field $\Phi$ with charge $q$, we use the gauge-covariant derivative $D_\mu\Phi=(\nabla_\mu-\ii qA_\mu)\Phi$, where $q$ is measured in units of the elementary charge.

We impose radial gauge and the horizon falloffs
\begin{equation}
A_r=0,\qquad A_u=\mathcal O(x),\qquad A_{z,\bar z}=\mathcal O(1).
\end{equation}
The residual gauge transformations are parametrized by an arbitrary smooth function $\eps(z,\bar z)$. We use the index $A$ for the $(z,\bar z)$ coordinates. For smooth angular fields and currents, we use the near-horizon expansions
\begin{equation}
\begin{aligned}
A_A(u,r,z,\bar z)&\sim\sum_{m=0}^{\infty}A_A^m(u,z,\bar z)x^m,\\
j_A(u,r,z,\bar z)&\sim\sum_{m=0}^{\infty}j_A^m(u,z,\bar z)x^m.
\end{aligned}\label{eq-series}
\end{equation}
Thus $A_A^0$ denotes the angular potential on the horizon. We separately define the normalized electric datum and matter flux by
\begin{equation}
A_u^0=\frac{\rh^2}{c_h}\left.\partial_rA_u\right|_{r=\rh},\qquad j_u^0=\rh^2\left.j_u\right|_{r=\rh}.
\end{equation}
In particular, $A_u^0$ is an electric datum, not the horizon value of $A_u$, which vanishes under the stated falloff. We allow smooth finite currents with $j_r=\mathcal O(1)$ and $j_u=\mathcal O(1)$. No vanishing condition on $j_r$ at the horizon is required. These falloffs allow a finite flux through the horizon, since $j^r=-j_u/c_h$ there. The factor $c_h$ is compensated by the null surface measure.

The Maxwell equations determine how these data extend away from the horizon. The $u$ component is a radial constraint whose integration fixes $A_u$ in terms of the angular fields, the bulk current, and the integration function $A_u^0$. Its integral form, together with the radial continuation of the conserved current, is given in Appendix~\ref{app-normalization}. The angular components are
\begin{equation}
\begin{aligned}
g^2c j_z={}&-2\partial_u\partial_rA_z +\partial_r\left(\frac{f}{c}\partial_rA_z\right) +\partial_r\partial_zA_u\\
&+\frac{c}{r^2}\partial_z\left[\gamma_{z\bar z}^{-1} (\partial_{\bar z}A_z-\partial_zA_{\bar z})\right],
\end{aligned}
\end{equation}
\begin{equation}
\begin{aligned}
g^2c j_{\bar z}={}&-2\partial_u\partial_rA_{\bar z} +\partial_r\left(\frac{f}{c}\partial_rA_{\bar z}\right) +\partial_r\partial_{\bar z}A_u\\
&+\frac{c}{r^2}\partial_{\bar z}\left[\gamma_{z\bar z}^{-1} (\partial_zA_{\bar z}-\partial_{\bar z}A_z)\right].
\end{aligned}
\end{equation}
Substituting Eq.~\eqref{eq-series}, the leading $z$ component gives
\begin{equation}
\begin{aligned}
2\partial_uA_z^1={}&\frac{f'(\rh)}{c_h}A_z^1 +\frac{c_h}{\rh^2}\partial_zA_u^0-g^2c_hj_z^0\\
&+\frac{c_h}{\rh^2}\partial_z\left[\gamma_{z\bar z}^{-1} (\partial_{\bar z}A_z^0-\partial_zA_{\bar z}^0)\right].
\end{aligned}
\end{equation}
The conjugate equation governs $A_{\bar z}^1$. Higher orders similarly determine the $u$ evolution of the subleading radial coefficients once their values on a reference cut are specified. The leading fields $A_z^0$ and $A_{\bar z}^0$ remain freely specified radiative data.

The remaining $r$ component, evaluated at the horizon, gives the supplementary constraint
\begin{equation}
g^2j_u^0=-\gamma_{z\bar z}^{-1}\partial_u(\partial_zA_{\bar z}^0+\partial_{\bar z}A_z^0)-\partial_uA_u^0. \label{eq-constraint}
\end{equation}
With the electric normalization above, this equation contains no explicit dependence on $f$ or $h$. The cancellation follows from $\sqrt{-g}=c(r)r^2\gamma_{z\bar z}$ and $g^{ur}=-1/c(r)$ and holds for a general smooth $c(r)$. Equation~\eqref{eq-constraint} determines the evolution of $A_u^0$ from the radiative and matter fluxes, leaving its value at one end of the horizon as independent electric data. We retain this endpoint contribution when constructing the large-gauge charge below.

\subsection{Charge generators and the Ward identity}

For a smooth gauge parameter $\eps(z,\bar z)$, the covariant surface charge can be written as
\begin{equation}
\mathcal Q_\eps =g^{-2}\int\eps\star F
\end{equation}
evaluated on a horizon sphere \cite{Iyer:1994ys,Barnich:2001jy}. In the retarded reference orientation, it takes the form
\begin{equation}
\mathcal Q_\eps(u)=\frac{1}{g^2}\int\dd^2z~\gamma_{z\bar z}\eps A_u^0(u,z,\bar z).
\end{equation}
We assume that the fields have well-defined limits as $u\to\pm\infty$ and that the integrated fluxes are finite. Integrating Eq.~\eqref{eq-constraint} along the branch gives
\begin{equation}
Q_\eps\equiv\mathcal Q_\eps(-\infty) =Q_{\mathrm{soft}}+Q_{\mathrm{hard}}+Q_{{\rm end},\eps},
\end{equation}
where
\begin{eqnarray}
Q_{\mathrm{soft}}&=&\frac{1}{g^2}\int_{\cH}\dd u\dd^2z\eps\partial_u (\partial_zA_{\bar z}^0+\partial_{\bar z}A_z^0)\,,\\
Q_{\mathrm{hard}}&=&\int_{\cH}\dd u\dd^2z\gamma_{z\bar z}\eps j_u^0\,,\\
Q_{{\rm end},\eps}&=&\mathcal Q_\eps(+\infty)\,.
\end{eqnarray}
The soft and hard terms describe the flux through the horizon, while $Q_{{\rm end},\eps}$ retains the electric datum that the flux does not determine.

To establish that $Q_\eps$ generates the large gauge transformation, we specify the endpoint conditions in the phase space \cite{Donnelly:2016auv,Chandrasekaran:2018aop,Harlow:2019yfa,Ball:2024yze}. We work at fixed $A_u^0(+\infty,z,\bar z)$ and retain its contribution to the charge. For the electric soft theorem, the endpoint connections satisfy
\begin{equation}
F_{z\bar z}|_{u=\pm\infty}=0, \qquad A_A^0|_{u=\pm\infty}=\partial_A\phi_\pm. \label{eq-flat-endpoints}
\end{equation}
The phases $\phi_\pm$ remain free to vary, so the endpoint conditions preserve the large gauge transformations.

For canonically normalized, minimally coupled complex scalar matter, with $\Phi^0=\Phi|_{r=\rh}$, the symplectic form in this fixed electric sector is
\begin{equation}
\begin{aligned}
\Omega_{\cH}={}&\frac{1}{g^2}\int_{\cH}\dd u\dd^2z \left[\delta(\partial_uA_z^0)\wedge\delta A_{\bar z}^0 +\delta(\partial_uA_{\bar z}^0)\wedge\delta A_z^0\right]\\
&+\int_{\cH}\dd u\dd^2z\gamma_{z\bar z}\rh^2 \left[\delta(\partial_u\bar\Phi^0)\wedge\delta\Phi^0 +\delta(\partial_u\Phi^0)\wedge\delta\bar\Phi^0\right].
\end{aligned}\label{eq-symplectic}
\end{equation}
The photon field includes its endpoint modes, and the matter wavepackets vanish at $u=\pm\infty$. Contracting this form with $\delta_\eps A_A^0=\partial_A\eps$ and $\delta_\eps\Phi^0=\ii q\eps\Phi^0$ gives
\begin{equation}
\Omega_{\cH}(\delta_\eps,\delta) =\delta Q_{\mathrm{soft}}+\delta Q_{\mathrm{hard}} =\delta Q_\eps,
\end{equation}
where the last equality uses $\delta Q_{{\rm end},\eps}=0$ within the chosen sector. Thus the complete charge generates
\begin{equation}
[Q_\eps,A_z^0]=\ii\partial_z\eps, \qquad [Q_\eps,\Phi_k^0]=-q_k\eps\Phi_k^0. \label{eq-gauge-action}
\end{equation}
The endpoint term fixes the sector-dependent value of the generator and must be retained when comparing different electric sectors.

The same symplectic form fixes the photon normalization. Writing $A_A^{\mathrm{rad}}$ for the radiative part, its interior bracket is
\begin{equation}
[A_z^{\mathrm{rad}}(u,z),A_{\bar w}^{\mathrm{rad}}(u',w)]_D =-\frac{\ii g^2}{4}\operatorname{sgn}(u-u')\delta^2(z-w). \label{eq-radiative-bracket}
\end{equation}
The endpoint brackets are fixed by Eq.~\eqref{eq-flat-endpoints} and the continuity prescription of Ref.~\cite{He:2014cra}. Together they reproduce the action on the full field in Eq.~\eqref{eq-gauge-action}. The explicit endpoint brackets and the matter normalization check are collected in Appendix~\ref{app-normalization}.

We now compare the past and future horizon branches, with generators $Q_\eps^-$ and $Q_\eps^+$. Their physical orientations are inherited from the static region. The outward direction points toward decreasing areal radius at a black hole horizon and increasing areal radius at a cosmological horizon, giving the relative sign
\begin{equation}
\eta=\begin{cases} +1,&\text{at a black hole horizon},\\
-1,&\text{at a cosmological horizon}.
\end{cases}
\end{equation}
This sign enters when the reference component expressions are assigned to the physical boundary. The two branches use the retarded and advanced charts described in Appendix~\ref{sec-coordinates}, whose roles are interchanged at a cosmological horizon.

The branches meet at the bifurcation sphere $B$. We choose the antipodal relabelling $\mathcal A(z,\bar z)=(-1/\bar z,-1/z)$ and impose
\begin{equation}
\eps^+=\mathcal A^*\eps^-, \qquad \left.\star F^+\right|_B=-\mathcal A^*\left(\left.\star F^-\right|_B\right), \qquad \phi_B^+=\mathcal A^*\phi_B^-. \label{eq-matching}
\end{equation}
Here the superscripts distinguish the two horizon branches, and the matching is expressed in common reference orientations. The phases $\phi_B^\pm$ are the endpoint phases restricted to $B$ on the two branches; these superscripts label branches, whereas the subscripts on $\phi_\pm$ label the limits $u\to\pm\infty$ on one branch. The minus sign accounts for the orientation reversal under the angular relabelling, with the coordinate and orientation conventions given in Appendix~\ref{sec-coordinates}. This prescription matches the endpoint data and gauge parameter while leaving the radiative fields on the two branches independent. For the soft theorem, the matched endpoint electric data are fixed, while the endpoint phases remain free to transform.

Charge conservation relates the horizon charges to the charge exchanged through the other boundaries. Let $\cS$ denote the full evolution operator, with the initial and final states specified on all boundary components. Charge conservation gives
\begin{equation}
\vev{\mathrm{out}|Q_\eps^+\cS-\cS Q_\eps^-|\mathrm{in}} =-\Delta_\eps. \label{eq-ward}
\end{equation}
Here $Q_\eps^\pm$ are the horizon charges, and $\Delta_\eps$ denotes the analogous matrix element for the other boundaries, with the outgoing charge acting to the left of $\cS$ and the incoming charge to its right. For the Schwarzschild exterior, this contribution comes from $\cI^+$ and $\cI^-$. For a region between two horizons, it comes from the other horizon.

The horizon Ward identity is homogeneous when $\Delta_\eps=0$. A sufficient condition is that the initial and final states on the other boundaries have equal, possibly nonzero, charge eigenvalues. We choose a fixed electric sector in which the endpoint contributions satisfy
\begin{equation}
\vev{\mathrm{out}|Q_{{\rm end},\eps}^{+}\cS-\cS Q_{{\rm end},\eps}^{-}|\mathrm{in}}=0.
\label{eq-endpoint-sector}
\end{equation}
This sector condition is independent of $\Delta_\eps=0$. Together they leave a relation between the soft and hard charges. This relation is the starting point for the horizon soft theorem.

\section{Horizon soft theorem}
\label{sec-soft-theorem}

The horizon Ward identity relates a zero-frequency photon insertion to the charges carried by the hard particles. In this section, we first derive the angular relation and then obtain its local momentum form with the same emitted mode. This gives a common soft theorem for the backgrounds considered below, subject to the stated identification of horizon and momentum states.


We work in the matched, fixed endpoint electric sector and take $\Delta_\eps=0$. The endpoint contributions then cancel, and Eq.~\eqref{eq-ward} reduces to
\begin{equation}
\begin{aligned}
&\vev{\mathrm{out}|Q_{\mathrm{soft}}^+\cS-\cS Q_{\mathrm{soft}}^-|\mathrm{in}}\\
&\qquad=-\vev{\mathrm{out}|Q_{\mathrm{hard}}^+\cS-\cS Q_{\mathrm{hard}}^-|\mathrm{in}}.
\end{aligned}
\label{eq-soft-hard-ward}
\end{equation}

To evaluate the soft contribution, choose $\eps=(w-z)^{-1}$, understood as a distributional limit of smooth gauge parameters on the sphere. Using $\partial_{\bar z}(w-z)^{-1}=-2\pi\delta^2(z-w)$ and the flat endpoint condition in Eq.~\eqref{eq-flat-endpoints}, the soft charge in the retarded reference orientation becomes
\begin{equation}
Q_{\mathrm{soft}}(w) =\frac{4\pi}{g^2}\int_{\cH}\dd u\,\partial_u A_w^0 =\frac{4\pi}{g^2} \left[A_w^0\right]_{u=-\infty}^{u=+\infty}.
\label{eq-soft-endpoint}
\end{equation}
The soft insertion is therefore determined by the change in the angular potential between the two endpoints.

To express this change in terms of photon operators, expand the nonzero-frequency radiative field as
\begin{equation}
A_z^{\mathrm{rad}}(u,z,\bar z) =-\frac{\ii\sqrt2g}{8\pi^2(1+z\bar z)} \int_0^\infty\dd\omega \left[b_+(\omega,z)e^{-\ii\omega u} -b_-^\dagger(\omega,z)e^{\ii\omega u}\right].
\label{eq-horizon-mode}
\end{equation}
Here $\omega$ is conjugate to $u$, and $z$ denotes the angular point $(z,\bar z)$. The normalization follows from Eq.~\eqref{eq-symplectic} and gives
\begin{equation}
[b_\alpha(\omega,z),b_\beta^\dagger(\omega',w)] =\frac{2(2\pi)^3}{\omega} \delta_{\alpha\beta}\delta(\omega-\omega') \delta_\Omega^2(z,w),
\label{eq-mode-bracket}
\end{equation}
with $\delta_\Omega^2$ normalized using the sphere measure $\dd\Omega$. The endpoint phases remain part of the phase space, and their difference in Eq.~\eqref{eq-soft-endpoint} is represented by the zero-frequency limit of these modes. All angular kernels and soft insertions are understood after wavepacket smearing.

Substituting the mode expansion into Eq.~\eqref{eq-soft-endpoint} yields
\begin{equation}
Q_{\mathrm{soft}}(w) =-\frac{1}{\sqrt2g(1+w\bar w)} \lim_{\omega\to0^+}\omega \left[b_+(\omega,w)+b_-^\dagger(\omega,w)\right].
\label{eq-soft-charge}
\end{equation}
The frequency integral runs over $\omega>0$, so the delta function produced by the $u$ integral contributes half its weight at $\omega=0$. The distributional normalization is checked in Appendix~\ref{app-normalization}.

The hard contribution follows from the gauge action in Eq.~\eqref{eq-gauge-action}. For particles crossing the horizon at angular positions $w_k$, it gives
\begin{equation}
\begin{aligned}
&\vev{\mathrm{out}|Q_{\mathrm{hard}}^+\cS-\cS Q_{\mathrm{hard}}^-|\mathrm{in}}\\
&\qquad= \left[ \sum_{l\in\mathrm{out}}\frac{q_l}{w-w_l} -\sum_{k\in\mathrm{in}}\frac{q_k}{w-w_k} \right] \vev{\mathrm{out}|\cS|\mathrm{in}}.
\end{aligned}
\label{eq-hard-charge-action}
\end{equation}

For hard states containing no photons in the relevant soft frequency band, $\langle\mathrm{out}|b_-^{{\rm out}\dagger}=0$ and $b_+^{\rm in}|\mathrm{in}\rangle=0$. Inserting the soft and hard contributions into Eq.~\eqref{eq-soft-hard-ward}, with the physical horizon orientations included through $\eta$, gives
\begin{equation}
\begin{aligned}
&\bra{\mathrm{out}} b_+(\omega,w)\cS-\cS b_-^\dagger(\omega,w) \ket{\mathrm{in}}\\
&\qquad\simeq 2\eta g\frac{1+w\bar w}{\sqrt2\omega} \left[ \sum_{l\in\mathrm{out}}\frac{q_l}{w-w_l} -\sum_{k\in\mathrm{in}}\frac{q_k}{w-w_k} \right] \vev{\mathrm{out}|\cS|\mathrm{in}}.
\end{aligned}
\label{eq-soft-theorem}
\end{equation}
Here $\simeq$ denotes equality of the leading soft terms. The operators to the left and right of $\cS$ belong to the outgoing and incoming horizon branches, respectively. For states containing a soft photon cloud, the two operator terms removed by the vacuum conditions must also be retained.

The soft factor obtained in Eq.~\eqref{eq-soft-theorem} will be abbreviated as
\begin{equation}
S(w)=\frac{1+w\bar w}{\sqrt2\omega} \left[ \sum_{l\in\mathrm{out}}\frac{q_l}{w-w_l} -\sum_{k\in\mathrm{in}}\frac{q_k}{w-w_k} \right].
\label{eq-angular-factor}
\end{equation}
The factor $S(w)$ includes the soft frequency pole; its frequency dependence is implicit. The crossed soft insertion in Eq.~\eqref{eq-soft-theorem} thus has coefficient $2\eta gS(w)$. If soft crossing holds for the chosen evolution problem, the incoming insertion equals minus the outgoing one at leading order. The single-emission amplitude then factorizes as
\begin{equation}
\vev{\mathrm{out}|b_+(\omega,w)\cS|\mathrm{in}} =\eta gS(w) \vev{\mathrm{out}|\cS|\mathrm{in}} +o(\omega^{-1}).
\label{eq-emission}
\end{equation}
This gives the familiar leading soft pole for a normalized horizon photon. Its residue depends on the charges and horizon crossing positions of the hard particles.


To compare the horizon soft theorem across explicit backgrounds, we express the single-photon emission relation in isotropic coordinates. The coordinate transformation in Appendix~\ref{sec-coordinates},
\begin{equation}
\frac{\dd\rho}{\rho}=\frac{\sqrt{h(r)}}{r}\dd r,
\end{equation}
puts the metric in the form
\begin{equation}
\dd s^2=-A^2(\rho)\dd t^2 +B^2(\rho)\left(\dd\rho^2+\rho^2\dd\Omega_2^2\right),
\end{equation}
with
\begin{equation}
A(\rho)=\sqrt{f(r(\rho))}, \qquad B(\rho)=\frac{r(\rho)}{\rho}.
\end{equation}
Let $\rho_H$ denote the horizon position and $R=|\rho-\rho_H|$ the positive coordinate distance into the static region. The frequency $\omega$ is conjugate to the Killing time $t$ and, at fixed radius, also to the null coordinate along the branch.

For a fixed angular wavepacket, we define the normalized isotropic coefficient $a_\alpha$ through
\begin{equation}
b_\alpha(\omega,w;R) =\frac{4\ii\omega}{g\mathcal K(R)} a_\alpha(\omega,w;R), \qquad \mathcal K(R)=\frac{A(R)}{B^4(R)\rho(R)}.
\label{eq-mode-map}
\end{equation}
This convention and its relation to a local Fourier representation are discussed in Appendix~\ref{app-normalization}. We choose the leading soft residue of $b_\alpha(\omega,w;R)$ to equal that of the normalized horizon mode. This specifies the stretched-horizon representative used below.

For a nondegenerate horizon, $A(R)=\kappa B_H R+\mathcal O(R^2)$, where $B_H=\rh/\rho_H$ and $\kappa=|f'(\rh)|/(2c_h)$. Thus the conversion factor behaves as
\begin{equation}
\mathcal K(R) =\frac{\kappa\rho_H^2}{\rh^3}R+\mathcal O(R^2).
\label{eq-isotropic-leading-coefficient}
\end{equation}
Substituting Eq.~\eqref{eq-mode-map} into the single-emission relation \eqref{eq-emission} gives
\begin{equation}
\vev{\mathrm{out}|a_+(\omega,w;R)\cS|\mathrm{in}}\simeq \eta g^2\frac{\mathcal K(R)}{4\ii\omega} S(w) \vev{\mathrm{out}|\cS|\mathrm{in}}\,.
\label{eq-local-theorem}
\end{equation}
Here $\simeq$ denotes the leading soft behavior at fixed $R>0$, followed by the small-$R$ expansion. The angular soft factor retains its familiar form, while $\mathcal K(R)$ accounts for the conversion between normalized horizon modes and their isotropic representation.

The angular soft factor can be expressed through local momentum contractions, following Ref.~\cite{Cheng:2022xyr}. For this representation, we assume that the external horizon states have been matched to massless momentum modes, with $w_k$ denoting their momentum directions. For an inward radial trajectory, this direction is antipodal to the outward radial vector at the crossing point. The corresponding angular relabelling and polarization convention must therefore be included in the identification of states on each branch. The same matching is assumed for the emitted photon; horizon crossing positions alone do not determine momentum directions.

At fixed $R>0$, write the soft null momentum and its positive-helicity polarization in isotropic Cartesian components as
\begin{equation}
\begin{aligned}
p_\mu &=\frac{\omega}{1+w\bar w} \left( -(1+w\bar w), \frac BA(w+\bar w), -\frac{\ii B}{A}(w-\bar w), \frac BA(1-w\bar w) \right),\\
\varepsilon^{+\mu} &=\frac1{\sqrt2 B} \left( \frac BA\bar w,1,-\ii,-\bar w \right).
\end{aligned}
\label{eq-local-momentum-polarization}
\end{equation}
Here $A$ and $B$ are evaluated at the same radius, and $p_\mu p^\mu=p_\mu\varepsilon^{+\mu}=0$. A hard null covector $k_{k\mu}$ has the same parametrization with $\omega$ replaced by its Killing energy $E_k$ and $w$ by $w_k$. Direct contraction gives
\begin{equation}
\begin{aligned}
k_{k\mu}\varepsilon^{+\mu} &=-\frac{\sqrt2 E_k}{A} \frac{\bar w-\bar w_k}{1+w_k\bar w_k},\\
k_{k\nu}p^\nu &=-\frac{2E_k\omega}{A^2} \frac{(w-w_k)(\bar w-\bar w_k)} {(1+w\bar w)(1+w_k\bar w_k)}.
\end{aligned}
\end{equation}
The hard energy cancels from their ratio, yielding
\begin{equation}
\frac{k_{k\mu}\varepsilon^{+\mu}}{k_{k\nu}p^\nu} =\frac{A(R)}{\omega} \frac{1+w\bar w}{\sqrt2(w-w_k)}.
\end{equation}
Consequently, the angular factor obeys
\begin{equation}
\sum_{l\in\mathrm{out}}q_l \frac{k_{l\mu}\varepsilon^{+\mu}}{k_{l\nu}p^\nu} -\sum_{k\in\mathrm{in}}q_k \frac{k_{k\mu}\varepsilon^{+\mu}}{k_{k\nu}p^\nu} =
A(R)\frac{1+w\bar w}{\sqrt2\omega} \left[ \sum_{l\in\mathrm{out}}\frac{q_l}{w-w_l} -\sum_{k\in\mathrm{in}}\frac{q_k}{w-w_k} \right] =A(R)S(w).
\label{eq-momentum-map}
\end{equation}
Substituting this identity into Eq.~\eqref{eq-local-theorem} gives the momentum-space representation
\begin{equation}
\begin{aligned}
&\vev{\mathrm{out}|a_+(\omega,w;R)\cS|\mathrm{in}}\\
&\quad\simeq \frac{\eta g^2\mathcal K(R)}{4\ii\omega A(R)} \left[ \sum_{l\in\mathrm{out}}q_l \frac{k_{l\mu}\varepsilon^{+\mu}}{k_{l\nu}p^\nu} -\sum_{k\in\mathrm{in}}q_k \frac{k_{k\mu}\varepsilon^{+\mu}}{k_{k\nu}p^\nu} \right] \vev{\mathrm{out}|\cS|\mathrm{in}}.
\end{aligned}
\label{eq-momentum-theorem-finite-R}
\end{equation}
The soft limit and radial expansion have the same ordering as in Eq.~\eqref{eq-local-theorem}. The geometric prefactor satisfies
\begin{equation}
\frac{\mathcal K(R)}{A(R)} =\frac1{B^4(R)\rho(R)}=\frac{\rho(R)^3}{r(R)^4} =C+\mathcal O(R).
\label{eq-momentum-coefficient}
\end{equation}
Here
\begin{equation}
C=\frac{\rho_H^3}{\rh^4}.
\end{equation}
Defining
\begin{equation}
S(p)=\left[ \sum_{l\in\mathrm{out}}q_l \frac{k_{l\mu}\varepsilon^{+\mu}}{k_{l\nu}p^\nu} -\sum_{k\in\mathrm{in}}q_k \frac{k_{k\mu}\varepsilon^{+\mu}}{k_{k\nu}p^\nu} \right]
\end{equation}
we can write the momentum-space soft theorem at leading radial order as
\begin{equation}
\vev{\mathrm{out}|a_+(\omega,w;R)\cS|\mathrm{in}}\simeq \eta g^2\frac{C}{4\ii\omega} S(p) \vev{\mathrm{out}|\cS|\mathrm{in}}.
\label{eq-momentum-theorem}
\end{equation}
Eq.~\eqref{eq-momentum-theorem} is the local momentum form of the same theorem, with the normalization of $a_+$ unchanged. Its leading geometric coefficient is $C$, and its prefactor is $\eta g^2 C/(4\ii\omega)$. The dependence of $S(p)=A(R)S(w)$ on $R$ and $\omega$ is implicit. Equation~\eqref{eq-momentum-theorem-finite-R} retains the full radial coefficient; Eq.~\eqref{eq-momentum-theorem} keeps its leading term after the soft limit.

In the homogeneous sector considered here, charge conservation gives $\sum_{\mathrm{out}}q_l-\sum_{\mathrm{in}}q_k=0$. The momentum factor is consequently invariant under $\varepsilon^{+\mu}\to\varepsilon^{+\mu}+\alpha p^\mu$. All contractions refer to local momenta at fixed $R$; connecting these modes to asymptotic particle states additionally requires propagation through the exterior geometry.

\section{Applications and extensions}
\label{sec-applications}

The preceding derivation applies to a general static, spherically symmetric nondegenerate horizon. In this section, familiar black hole and cosmological geometries provide explicit checks of the soft coefficients. We then consider backgrounds with independent metric functions, before extending the construction to Yang-Mills theory.

\subsection{Black hole and cosmological horizons}
\label{sec-backgrounds}

We consider familiar black hole and cosmological geometries here, which provide explicit checks of the soft theorems. All the examples in this subsection satisfy $h=f^{-1}$. Their normalized horizon modes obey Eq.~\eqref{eq-soft-theorem}. For each background, we evaluate $\mathcal K(R)$ and its momentum counterpart $\mathcal K(R)/A(R)=C+\mathcal O(R)$, which enter Eqs.~\eqref{eq-local-theorem} and \eqref{eq-momentum-theorem-finite-R} with the same photon normalization. The momentum form uses the local state matching specified above. The Schwarzschild and dS cases also allow comparison with earlier horizon soft theorems.

We work with matched, fixed endpoint electric data and impose $\Delta_\eps=0$ for the remaining boundaries. The single-emission formulas additionally assume the crossing relation used in Eq.~\eqref{eq-emission}. Throughout, the soft limit is taken at fixed $R>0$ before the stretched horizon is removed.

The coordinate coefficients follow directly from the horizon data. Let $\rho_H$ denote the isotropic horizon radius and define
\begin{equation}
R=\begin{cases} \rho-\rho_H, & \text{at a black hole horizon},\\
\rho_H-\rho, & \text{at a cosmological horizon}.
\end{cases}
\end{equation}
Thus $R>0$ on the static side in both cases. Using $\kappa=|f'(\rh)|/(2c_h)$, the isotropic transformation gives
\begin{equation}
\begin{aligned}
r-\rh&=\frac{f'(\rh)\rh^2}{4c_h^2\rho_H^2}R^2+\mathcal O(R^3),\\
A(R)&=\frac{\kappa\rh}{\rho_H}R+\mathcal O(R^2),\\
B(R)&=\frac{\rh}{\rho_H}+\mathcal O(R).
\end{aligned}
\label{eq-isotropic}
\end{equation}
Consequently
\begin{equation}
\frac{\mathcal K(R)}{4\ii\omega}=\frac{\kappa\rho_H^2}{4\ii\omega\rh^3}R+\mathcal O(R^2/\omega).
\label{eq-coefficients}
\end{equation}
Here $\mathcal K(R)>0$ on the static side for either type of horizon; the relative orientation is entirely carried by $\eta$.

Using the mode dictionary in Eq.~\eqref{eq-mode-map}, the single-emission relation can be represented in the isotropic frame as
\begin{equation}
\vev{\mathrm{out}|a_+(\omega,w;R)\cS|\mathrm{in}} \simeq\eta g^2\frac{\mathcal K(R)}{4\ii\omega}S(w) \vev{\mathrm{out}|\cS|\mathrm{in}}.
\label{eq-emission-coefficients}
\end{equation}
This is not an additional Ward identity, but the isotropic-coordinate representation of Eq.~\eqref{eq-emission}.

\paragraph{Schwarzschild black hole.}

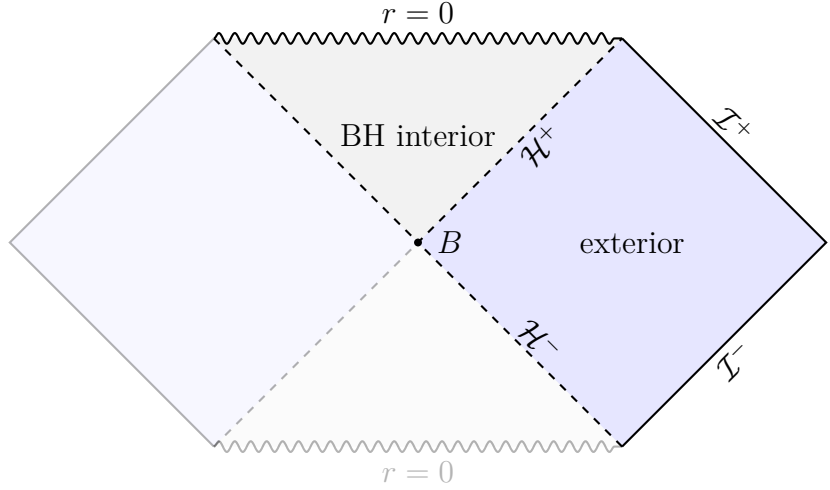
\begin{figure}[t]
\centering
\begin{tikzpicture}[scale=1.35,opacity=1]
\begin{scope}[causal context]
\foreach \s in {-1,1} { \path[causal exterior] (0,0)--({2*\s},2)--({4*\s},0)--({2*\s},-2)--cycle; }
\foreach \s in {-1,1} { \path[causal region] (-2,{2*\s})--(0,0)--(2,{2*\s})--cycle; }
\end{scope}
\path[causal exterior] (0,0)--(2,2)--(4,0)--(2,-2)--cycle;
\path[causal region] (-2,2)--(0,0)--(2,2)--cycle;
\begin{scope}[causal context]
\draw[causal boundary] (-2,-2)--(-4,0)--(-2,2);
\draw[causal horizon] (-2,-2)--(0,0);
\draw[causal singularity] (-2,-2)--(2,-2);
\node at (0,-2.25) {$r=0$};
\end{scope}
\draw[causal boundary] (2,-2)--(4,0)--(2,2);
\draw[causal horizon] (2,-2)--(0,0)--(2,2);
\draw[causal horizon] (-2,2)--(0,0);
\draw[causal singularity] (-2,2)--(2,2);
\node at (2.1,0) {exterior};
\node at (0,1.05) {BH interior};
\node[rotate=45] at (1.2,0.95) {$\cH^+$};
\node[rotate=315] at (1.2,-0.95) {$\cH^-$};
\node[rotate=315] at (3.1,1.15) {$\cI^+$};
\node[rotate=45] at (3.1,-1.15) {$\cI^-$};
\node at (0,2.25) {$r=0$};
\fill (0,0) circle (1.1pt);
\node[right=3pt] at (0,0) {$B$};
\end{tikzpicture}
\caption{Maximally extended Schwarzschild geometry. The exterior regions are in blue, and the interior regions are in light gray.  The horizon branches $\cH^\pm$ meet at the bifurcation sphere $B$.}
\label{fig-schwarzschild}
\end{figure}

We begin with Schwarzschild, for which the isotropic transformation can be evaluated explicitly. The metric and horizon data are
\begin{equation}
f(r)=1-\frac{2M}{r},\qquad h(r)=f(r)^{-1},\qquad \rh=2M,\qquad \kappa=\frac1{4M}.
\end{equation}
The selected exterior and its boundaries are shown in Fig.~\ref{fig-schwarzschild}. Fixing the isotropic scale by $\rho/r\to1$ at spatial infinity gives
\begin{equation}
r=\rho\left(1+\frac{M}{2\rho}\right)^2,\qquad \rho_H=\frac M2.
\end{equation}
With $R=\rho-M/2$, the metric functions become
\begin{equation}
A(R)=\frac{R}{R+M},\qquad B(R)=\frac{4(R+M)^2}{(2R+M)^2}.
\end{equation}
Hence $A(R)=R/M+\mathcal O(R^2)$ and $B_H=4$, which give
\begin{equation}
\frac{\mathcal K(R)}{4\ii\omega}=\frac{R}{512\ii\omega M^2}+\mathcal O(R^2/\omega).
\end{equation}
This coefficient agrees with Eq.~\eqref{eq-coefficients}.

For a black hole horizon $\eta=+1$. The Ward relation itself remains Eq.~\eqref{eq-soft-theorem}, while its single-emission form is Eq.~\eqref{eq-emission}. Using the Schwarzschild value of $\mathcal K(R)$, the isotropic representation becomes
\begin{equation}
\vev{\mathrm{out}|a_+(\omega,w;R)\cS|\mathrm{in}} \simeq\frac{g^2R}{512\ii\omega M^2}S(w) \vev{\mathrm{out}|\cS|\mathrm{in}}.
\label{eq-schwarzschild-emission}
\end{equation}
The angular pole structure and the $R/M^2$ dependence agree with the Schwarzschild horizon soft theorem obtained from large gauge symmetry in Ref.~\cite{Cheng:2022xyr}. The difference in presentation is that the Ward identity is written first in terms of the symplectically normalized horizon mode $b_\alpha$, while the Schwarzschild-dependent coefficient appears only when that mode is represented by $a_\alpha$ in the isotropic chart.

The momentum coefficient is $C=1/(128M)$, so Eq.~\eqref{eq-momentum-theorem} gives, at leading radial order
\begin{equation}
\vev{\mathrm{out}|a_+(\omega,w;R)\cS|\mathrm{in}} \simeq\frac{g^2}{512\ii\omega M}S(p)\vev{\mathrm{out}|\cS|\mathrm{in}}.
\label{eq-schwarzschild-momentum}
\end{equation}
For $g=1$, the prefactor agrees with the $1/(512\ii\omega M)$ coefficient in Ref.~\cite{Cheng:2022xyr}.

\paragraph{Schwarzschild-Anti de Sitter (S-AdS) black hole.}

S-AdS provides a useful non-asymptotically flat example in which the isotropic conversion can be worked out explicitly near the event horizon. For an AdS radius $\ell$, the metric functions are
\begin{equation}
f(r)=1-\frac{2M}{r}+\frac{r^2}{\ell^2},\qquad h(r)=f(r)^{-1}.
\end{equation}
Let $r_S$ denote the event-horizon radius. The condition $f(r_S)=0$ gives
\begin{equation}
2M=r_S\left(1+\frac{r_S^2}{\ell^2}\right),
\end{equation}
and the surface gravity is
\begin{equation}
\kappa=\frac{f'(r_S)}{2}=\frac{1+3r_S^2/\ell^2}{2r_S}.
\end{equation}

We can also determine the isotropic coefficient directly. Using the horizon condition to eliminate $M$, the metric function can be factorized as
\begin{equation}
f(r)=\frac{(r-r_S)(r^2+r_Sr+r_S^2+\ell^2)}{\ell^2r}.
\label{eq-sads-factor}
\end{equation}
The isotropic coordinate is defined by
\begin{equation}
\frac{\dd\rho}{\rho}=\frac{\dd r}{r\sqrt{f(r)}},
\end{equation}
or equivalently
\begin{equation}
\log\frac{\rho}{\rho_H}=\int_{r_S}^{r}\frac{\dd s}{s\sqrt{f(s)}}.
\label{eq-sads-isotropic}
\end{equation}
The full integral is not needed for the horizon soft theorem. What enters the mode conversion is only its near-horizon behavior. We therefore set $r=r_S+\delta$, with $\delta>0$ in the exterior region. Equation~\eqref{eq-sads-factor} gives
\begin{equation}
f(r_S+\delta)=\frac{1+3r_S^2/\ell^2}{r_S}\delta+\mathcal O(\delta^2)=2\kappa\delta+\mathcal O(\delta^2).
\label{eq-sads-near-f}
\end{equation}
Substituting this expansion into Eq.~\eqref{eq-sads-isotropic}, we obtain
\begin{equation}
\log\frac{\rho}{\rho_H}=\sqrt{\frac{2\delta}{\kappa r_S^2}}+\mathcal O(\delta^{3/2}).
\label{eq-sads-rho-expansion}
\end{equation}
Writing $R=\rho-\rho_H$ and expanding $\log(\rho/\rho_H)=R/\rho_H+\mathcal O(R^2)$ then gives
\begin{equation}
r-r_S=\frac{\kappa r_S^2}{2\rho_H^2}R^2+\mathcal O(R^3).
\label{eq-sads-r-expansion}
\end{equation}

The overall scale of the isotropic radius remains arbitrary. We fix it by choosing $\rho_H=r_S/4$, which makes the normalization continuous with the Schwarzschild convention used above. Equation~\eqref{eq-sads-r-expansion} then becomes
\begin{equation}
r-r_S=8\kappa R^2+\mathcal O(R^3)=\frac{4(1+3r_S^2/\ell^2)}{r_S}R^2+\mathcal O(R^3).
\end{equation}
The two isotropic metric functions consequently behave as
\begin{equation}
A(R)=\sqrt{f(r(R))}=4\kappa R+\mathcal O(R^2)=\frac{2(1+3r_S^2/\ell^2)}{r_S}R+\mathcal O(R^2),
\end{equation}
and
\begin{equation}
B(R)=\frac{r(R)}{\rho(R)}=4+\mathcal O(R).
\end{equation}
Therefore the conversion coefficient defined in Eq.~\eqref{eq-mode-map} is
\begin{equation}
\frac{\mathcal K(R)}{4\ii\omega}=\frac{1+3r_S^2/\ell^2}{128\ii\omega r_S^2}R+\mathcal O(R^2/\omega).
\label{eq-sads-K}
\end{equation}

With $\rho_H=r_S/4$, the momentum conversion is
\begin{equation}
\frac{\mathcal K(R)}{A(R)}=C+\mathcal O(R),\qquad C=\frac1{64r_S}.
\end{equation}
The leading momentum prefactor is $g^2/(256\ii\omega r_S)$, which reduces to the Schwarzschild value as $\ell\to\infty$ and $r_S\to2M$.

The Ward identity itself remains the universal horizon relation in Eq.~\eqref{eq-soft-theorem}. For the event horizon $\eta=+1$, and imposing the crossing relation for a single soft emission gives Eq.~\eqref{eq-emission}. Using Eq.~\eqref{eq-sads-K}, its isotropic representation becomes
\begin{equation}
\vev{\mathrm{out}|a_+(\omega,w;R)\cS|\mathrm{in}} \simeq\frac{g^2R}{128\ii\omega r_S^2}\left(1+\frac{3r_S^2}{\ell^2}\right) S(w)\vev{\mathrm{out}|\cS|\mathrm{in}}.
\label{eq-sads-emission}
\end{equation}
Thus the cosmological constant does not alter the normalized horizon Ward identity or its angular pole structure. Its explicit contribution to the angular soft prefactor enters through the conversion between the normalized horizon mode and its isotropic representation, here through the factor $1+3r_S^2/\ell^2$.

The Schwarzschild limit provides a direct check. As $\ell\to\infty$, one has $r_S\to2M$ and
\begin{equation}
\frac{\mathcal K(R)}{4\ii\omega}\quad \to \quad\frac{R}{512\ii\omega M^2}+\mathcal O(R^2/\omega),
\end{equation}
so Eq.~\eqref{eq-sads-emission} reduces precisely to the Schwarzschild result in Eq.~\eqref{eq-schwarzschild-emission}.

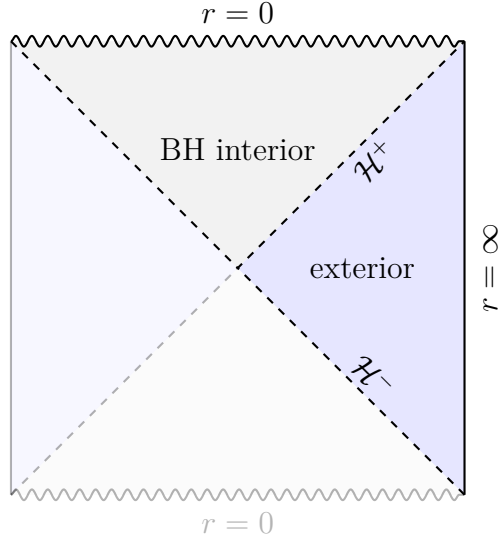
\begin{figure}[t]
\centering
\begin{tikzpicture}[scale=1.5,opacity=1]
\begin{scope}[causal context]
\foreach \s in {-1,1} { \path[causal exterior] (0,0)--({2*\s},2)--({2*\s},-2)--cycle; }
\foreach \s in {-1,1} { \path[causal region] (-2,{2*\s})--(0,0)--(2,{2*\s})--cycle; }
\end{scope}
\path[causal exterior] (0,0)--(2,2)--(2,-2)--cycle;
\path[causal region] (-2,2)--(0,0)--(2,2)--cycle;
\begin{scope}[causal context]
\draw[causal boundary] (-2,-2)--(-2,2);
\draw[causal horizon] (-2,-2)--(0,0);
\draw[causal singularity] (-2,-2)--(2,-2);
\node at (0,-2.25) {$r=0$};
\end{scope}
\draw[causal boundary] (2,-2)--(2,2);
\draw[causal horizon] (2,-2)--(0,0)--(2,2);
\draw[causal horizon] (-2,2)--(0,0);
\draw[causal singularity] (-2,2)--(2,2);
\node at (1.1,0) {exterior};
\node at (0,1.05) {BH interior};
\node[rotate=45] at (1.2,0.95) {$\cH^+$};
\node[rotate=315] at (1.2,-0.95) {$\cH^-$};
\node at (0,2.25) {$r=0$};
\node[rotate=90] at (2.23,0) {$r=\infty$};
\end{tikzpicture}
\caption{Maximally extended S-AdS geometry. The selected right exterior and future interior are emphasized together with all their boundaries. The solid timelike boundaries represent conformal infinity, and the wavy boundaries represent spacelike singularities.}
\label{fig-sads}
\end{figure}

There is one additional point specific to the AdS geometry. The exterior region has a timelike conformal boundary, shown in Fig.~\ref{fig-sads}, which enters the global charge balance even though it does not modify the local horizon calculation above. In the sector used here, we impose reflecting boundary conditions and fix the auxiliary charge so that $\Delta_\eps=0$. If flux through the timelike boundary were allowed, the corresponding boundary contribution would have to be retained in the Ward identity.

\paragraph{dS horizon.}

\begin{figure}[t]
\centering
\begin{tikzpicture}[scale=1.5,opacity=1]
\begin{scope}[causal context]
\foreach \s in {-1,1} { \path[causal exterior] (0,0)--({2*\s},2)--({2*\s},-2)--cycle; }
\foreach \s in {-1,1} { \path[causal region] (-2,{2*\s})--(0,0)--(2,{2*\s})--cycle; }
\end{scope}
\path[causal exterior] (0,0)--(2,2)--(2,-2)--cycle;
\path[causal region] (-2,2)--(0,0)--(2,2)--cycle;
\begin{scope}[causal context]
\draw[causal boundary] (-2,-2)--(-2,2);
\draw[causal boundary] (-2,-2)--(2,-2);
\draw[causal horizon] (-2,-2)--(0,0);
\end{scope}
\draw[causal boundary] (2,-2)--(2,2);
\draw[causal boundary] (-2,2)--(2,2);
\draw[causal horizon] (2,-2)--(0,0)--(2,2);
\draw[causal horizon] (-2,2)--(0,0);
\node[align=center] at (1.1,0) {static\\patch};
\node[rotate=45] at (1.2,0.95) {$\cH^+$};
\node[rotate=315] at (1.2,-0.95) {$\cH^-$};
\node at (0,2.25) {$r=\infty$};
\node[rotate=90] at (2.23,0) {$r=0$};
\end{tikzpicture}
\caption{Global dS geometry. The static patches are in blue, and the expanding and contracting cosmological regions are in light gray. The timelike boundaries at $r=0$ are regular centers.}
\label{fig-de-sitter}
\end{figure}
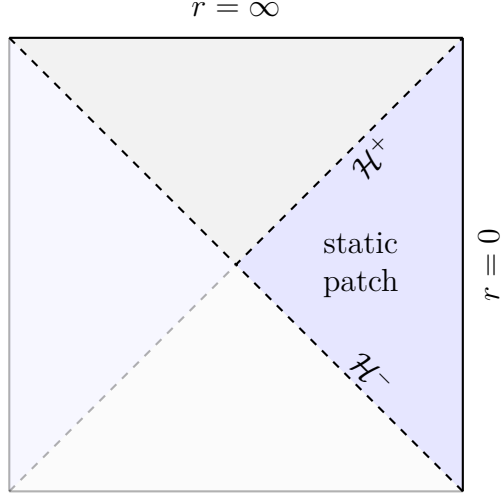

The dS static patch provides the corresponding check for a cosmological horizon. Its metric and horizon data are
\begin{equation}
f(r)=1-\frac{r^2}{\ell^2}, \qquad h(r)=f(r)^{-1}, \qquad \rh=\ell, \qquad \kappa=\frac1\ell.
\end{equation}
The calculation concerns the static patch bounded by $\cH^\pm$ in Fig.~\ref{fig-de-sitter}. Using the exact isotropic transformation
\begin{equation}
r=\frac{2\ell^2\rho}{\ell^2+\rho^2},
\end{equation}
we obtain
\begin{equation}
A(\rho)=\frac{\ell^2-\rho^2}{\ell^2+\rho^2}, \qquad B(\rho)=\frac{2\ell^2}{\ell^2+\rho^2}.
\end{equation}
Here $0\leq\rho<\rho_H=\ell$ covers the static patch, so $R=\ell-\rho$. Since $A(R)=R/\ell+\mathcal O(R^2)$ and $B_H=1$,
\begin{equation}
\frac{\mathcal K(R)}{4\ii\omega}=\frac{R}{4\ii\omega\ell^2}+\mathcal O(R^2/\omega),\qquad\eta=-1.
\end{equation}
The momentum conversion is
\begin{equation}
\frac{\mathcal K(R)}{A(R)}=C+\mathcal O(R),\qquad C=\frac1\ell.
\end{equation}
Together with $\eta=-1$, this gives the leading momentum prefactor $-g^2/(4\ii\omega\ell)$. Both representations carry the same cosmological horizon orientation sign. The angular conversion coefficient again agrees with Eq.~\eqref{eq-coefficients}. The single-emission coefficient multiplying $S(w)$ in Eq.~\eqref{eq-local-theorem} is therefore $\ii g^2R/(4\omega\ell^2)$ to leading order in $R$. The angular soft factor and the $R/\ell^2$ dependence agree with the cosmological horizon result of Ref.~\cite{Mao:2023rca}. That reference uses the crossed insertion and $R=\rho-\ell$; the radial, polarization and mode conventions must therefore be aligned before comparing overall coefficients. The relative sign between the black hole and cosmological cases follows from their boundary orientations, as encoded by $\eta$.

\paragraph{RN black hole.}

\begin{figure}[t]
\centering
\begin{tikzpicture}[scale=1.65,font=\small,opacity=1]
\begin{scope}[causal context]
\foreach \s in {-1,1} { \path[causal exterior] (0,0)--(\s,1)--({2*\s},0)--(\s,-1)--cycle; }
\foreach \t in {-1,1} { \path[causal region] (0,0)--(1,\t)--(1,{3*\t})--(-1,{3*\t})--(-1,\t)--cycle; }
\end{scope}
\path[causal exterior] (0,0)--(1,1)--(2,0)--(1,-1)--cycle;
\path[causal region] (0,0)--(1,1)--(0,2)--(-1,1)--cycle;
\path[causal region] (-1,1)--(0,2)--(-1,3)--cycle;
\begin{scope}[causal context]
\draw[causal boundary] (-1,-1)--(-2,0)--(-1,1);
\draw[causal horizon] (-1,-1)--(0,0);
\draw[causal horizon] (-1,-1)--(0,-2)--(1,-1);
\draw[causal horizon] (-1,-3)--(0,-2)--(1,-3);
\draw[causal horizon] (0,2)--(1,3);
\draw[causal singularity] (1,1)--(1,3);
\draw[causal singularity] (-1,-1)--(-1,-3);
\draw[causal singularity] (1,-1)--(1,-3);
\foreach \t in {-1,1} { \node at (0,{3.45*\t}) {$\vdots$}; }
\end{scope}
\draw[causal boundary] (1,-1)--(2,0)--(1,1);
\draw[causal horizon] (1,-1)--(0,0)--(1,1);
\draw[causal horizon] (0,0)--(-1,1);
\draw[causal horizon] (-1,1)--(0,2)--(1,1);
\draw[causal horizon] (0,2)--(-1,3);
\draw[causal singularity] (-1,1)--(-1,3);
\node at (1.1,0) {exterior};
\node[align=center] at (0,0.95) {between\\horizons};
\node[rotate=45,font=\footnotesize] at (0.65,0.40) {$\cH_+^+$};
\node[rotate=315,font=\footnotesize] at (0.65,-0.40) {$\cH_+^-$};
\node[rotate=45,font=\footnotesize] at (-0.53,1.66) {$r=r_-$};
\node[rotate=315] at (1.72,0.67) {$\cI^+$};
\node[rotate=45] at (1.72,-0.67) {$\cI^-$};
\node[rotate=90] at (-1.28,2.15) {$r=0$};
\end{tikzpicture}
\caption{A portion of the maximal analytic extension of nonextremal RN spacetime. The selected exterior, the future region between $r_+$ and $r_-$, and the adjoining upper left region $0<r<r_-$ are shown at full opacity, including their boundaries. The remaining extension is shown at reduced opacity, with ellipses indicating continuation to the past and future.}
\label{fig-rn}
\end{figure}
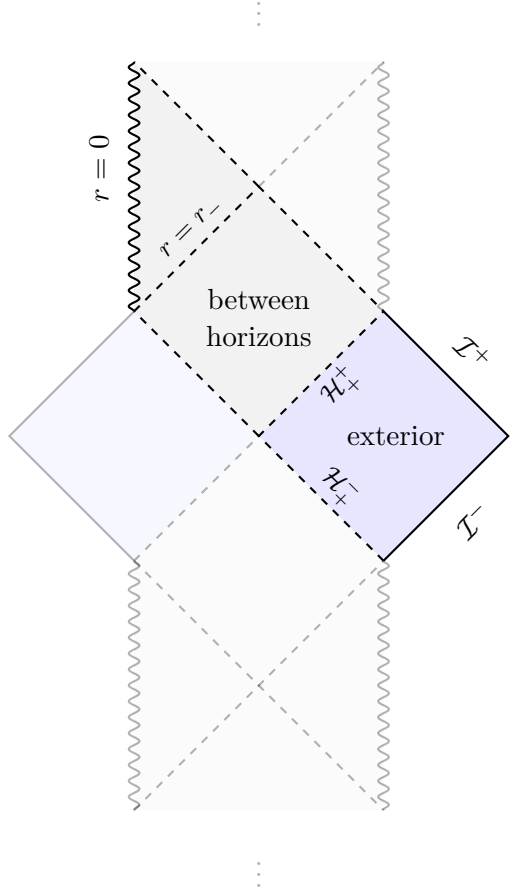

For the nonextremal charged black hole, define
\begin{equation}
\sigma=\sqrt{M^2-Q_{\mathrm{BH}}^2}>0, \qquad r_\pm=M\pm\sigma.
\end{equation}
The metric functions and isotropic transformation are
\begin{equation}
\begin{aligned}
f(r)&=\frac{(r-r_+)(r-r_-)}{r^2}, \qquad h(r)=f(r)^{-1},\\
r&=\rho+M+\frac{\sigma^2}{4\rho}.
\end{aligned}
\end{equation}
We apply the soft theorem to the outer event horizon shown in Fig.~\ref{fig-rn}, taking the Maxwell probe to be independent of the field supporting the background charge. The horizon data are
\begin{equation}
\rh=r_+, \qquad \rho_H=\frac{\sigma}{2}, \qquad \kappa=\frac{\sigma}{r_+^2}.
\end{equation}
Substitution into Eq.~\eqref{eq-coefficients} gives
\begin{equation}
\frac{\mathcal K(R)}{4\ii\omega}=\frac{\sigma^3 R}{16\ii\omega r_+^5}+\mathcal O(R^2/\omega),\qquad\eta=+1.
\end{equation}
For the outer horizon, the momentum conversion is
\begin{equation}
\frac{\mathcal K(R)}{A(R)}=C+\mathcal O(R),\qquad C=\frac{\sigma^3}{8r_+^4}.
\end{equation}
The leading momentum prefactor is $g^2\sigma^3/(32\ii\omega r_+^4)$. As $Q_{\mathrm{BH}}\to0$, it approaches $g^2/(512\ii\omega M)$.

The background charge thus changes the coordinate coefficients without introducing an additional factor into the normalized horizon soft theorem. When $Q_{\mathrm{BH}}\to0$, $\sigma\to M$ and $r_+\to2M$, so the coefficient reduces to the Schwarzschild values. The extremal case $\sigma=0$ lies outside the nondegenerate horizon expansion used here.

\paragraph{S-dS black hole.}

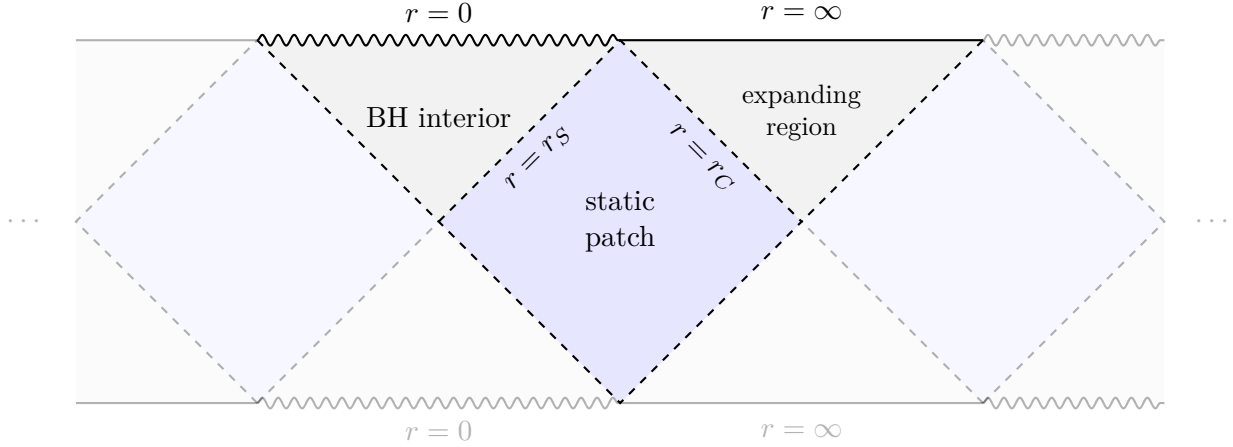
\begin{figure}[t]
\centering
\begin{tikzpicture}[scale=1.2,font=\small,opacity=1]
\begin{scope}[causal context]
\foreach \x in {0,4,8} { \path[causal exterior] ({\x-2},0)--(\x,2)--({\x+2},0)--(\x,-2)--cycle; }
\foreach \y in {-2,2} {
  \path[causal region] (-2,\y)--(-2,0)--(0,\y)--cycle;
  \path[causal region] (0,\y)--(2,0)--(4,\y)--cycle;
  \path[causal region] (4,\y)--(6,0)--(8,\y)--cycle;
  \path[causal region] (8,\y)--(10,0)--(10,\y)--cycle;
}
\end{scope}
\path[causal exterior] (2,0)--(4,2)--(6,0)--(4,-2)--cycle;
\path[causal region] (0,2)--(2,0)--(4,2)--cycle;
\path[causal region] (4,2)--(6,0)--(8,2)--cycle;
\begin{scope}[causal context]
\draw[causal horizon] (0,-2)--(-2,0)--(0,2);
\draw[causal horizon] (8,-2)--(10,0)--(8,2);
\draw[causal horizon] (0,-2)--(2,0);
\draw[causal horizon] (8,-2)--(6,0);
\draw[causal boundary] (-2,2)--(0,2);
\draw[causal boundary] (-2,-2)--(0,-2);
\draw[causal boundary] (4,-2)--(8,-2);
\draw[causal singularity] (0,-2)--(4,-2);
\draw[causal singularity] (8,2)--(10,2);
\draw[causal singularity] (8,-2)--(10,-2);
\node at (2,-2.3) {$r=0$};
\node at (6,-2.3) {$r=\infty$};
\node at (-2.55,0) {$\cdots$};
\node at (10.55,0) {$\cdots$};
\end{scope}
\draw[causal horizon] (2,0)--(4,2)--(6,0)--(4,-2)--cycle;
\draw[causal horizon] (0,2)--(2,0);
\draw[causal horizon] (8,2)--(6,0);
\draw[causal singularity] (0,2)--(4,2);
\draw[causal boundary] (4,2)--(8,2);
\node[align=center] at (4,0) {static\\patch};
\node at (2,1.15) {BH interior};
\node[align=center,font=\footnotesize] at (6,1.2) {expanding\\region};
\node[rotate=45] at (3.1,0.70) {$r=r_S$};
\node[rotate=315] at (4.9,0.70) {$r=r_C$};
\node at (2,2.3) {$r=0$};
\node at (6,2.3) {$r=\infty$};
\end{tikzpicture}
\caption{A portion of the maximally extended S-dS geometry. The central static region between $r_S$ and $r_C$ and its adjacent future nonstatic regions are shown at full opacity, including all their boundaries.}
\label{fig-sds}
\end{figure}

The Schwarzschild black hole with dS asymptotics contains both types of horizons and allows their coefficients to be checked within the same static region. The metric functions are
\begin{equation}
f(r)=1-\frac{2M}{r}-\frac{r^2}{\ell^2}, \qquad h(r)=f(r)^{-1}, \qquad \Lambda=\frac3{\ell^2}.
\end{equation}
For $0<M<\ell/(3\sqrt3)$, the two positive roots $r_S<r_C$ bound the static region in Fig.~\ref{fig-sds}. Their surface gravities are
\begin{equation}
\kappa_S=\frac{1-\Lambda r_S^2}{2r_S}, \qquad \kappa_C=\frac{\Lambda r_C^2-1}{2r_C}.
\end{equation}

To make contact with the preceding examples, choose $\rho_{H,S}=r_S/4$ near the black hole horizon and $\rho_{H,C}=r_C$ near the cosmological horizon. These choices refer to separately normalized local isotropic charts. Equation~\eqref{eq-coefficients} yields
\begin{equation}
\begin{aligned}
\frac{\mathcal K_S(R)}{4\ii\omega}&=\frac{1-\Lambda r_S^2}{128\ii\omega r_S^2}R+\mathcal O(R^2/\omega),\\
\frac{\mathcal K_C(R)}{4\ii\omega}&=\frac{\Lambda r_C^2-1}{8\ii\omega r_C^2}R+\mathcal O(R^2/\omega).
\end{aligned}
\end{equation}
In each chart, $R$ increases into the static region, with $\eta_S=+1$ and $\eta_C=-1$.

In the separately normalized isotropic charts, the momentum coefficients are
\begin{equation}
\begin{aligned}
\frac{\mathcal K_S(R)}{A_S(R)}&=C_S+\mathcal O(R),& \quad \text{with}\quad C_S&=\frac1{64r_S},\\
\frac{\mathcal K_C(R)}{A_C(R)}&=C_C+\mathcal O(R),& \quad \text{with}\quad C_C&=\frac1{r_C}.
\end{aligned}
\end{equation}
The leading momentum prefactors are $g^2/(256\ii\omega r_S)$ at the black hole horizon and $-g^2/(4\ii\omega r_C)$ at the cosmological horizon. They recover the Schwarzschild and dS coefficients in the respective limits below.

The black hole coefficients recover the Schwarzschild result as $\Lambda\to0$ at fixed $M$. The cosmological coefficients recover the dS result as $M\to0$ at fixed $\ell$. These limits check both the numerical factors and the relative orientation of the two horizon soft relations. The coincident-horizon limit is degenerate and is excluded from the present analysis. For a Ward identity associated with either horizon, charge exchanged through the other horizon contributes to $\Delta_\eps$.

\subsection{Non-trivial examples}
\label{sec-independent-backgrounds}

We now examine two backgrounds that test the role of Lorentz symmetry and the need for a constant product $f(r)h(r)$ in the horizon soft theorem.

\paragraph{A black hole in bumblebee gravity.}

The connection between soft photon theorems and Lorentz invariance motivates an examination of backgrounds with spontaneous Lorentz symmetry breaking. In the conventional scattering argument, Lorentz invariance for massless spin-one particles requires longitudinal polarizations to decouple from physical amplitudes, while the soft pole structure relates this requirement to charge conservation \cite{Weinberg:1964ew,Weinberg:1965nx}. The same leading soft relation can be understood as the Ward identity of angle-dependent large gauge transformations \cite{He:2014cra}. These perspectives raise a natural question for the horizon construction. Does Lorentz symmetry breaking in the gravitational background obstruct the large gauge transformations or modify the soft relation generated by their charges? Since internal gauge symmetry and Lorentz symmetry are distinct, this question must be examined through the probe action and its boundary conditions.

Bumblebee gravity provides a concrete setting for this question \cite{Casana:2017jkc}. The theory contains a vector field $B_\mu$ whose potential favors a nonzero vacuum expectation value. For the spacelike vacuum considered here, the action is
\begin{equation}
S_{\mathrm B} =\int\dd^4x\sqrt{-g} \left[ \frac{\mathcal R+\xi B^\mu B^\nu R_{\mu\nu}}{16\pi G} -\frac14 B_{\mu\nu}B^{\mu\nu} -V(B_\mu B^\mu-b^2) \right],
\end{equation}
where $\mathcal R$ is the Ricci scalar, $B_{\mu\nu}=\partial_\mu B_\nu-\partial_\nu B_\mu$, and $\xi$ controls the nonminimal coupling to curvature. At the minimum of the potential, $\langle B_\mu\rangle=b_\mu$ satisfies $b_\mu b^\mu=b^2>0$. This vacuum selects a preferred local direction and spontaneously breaks local Lorentz symmetry. The Schwarzschild-like solution uses a radial spacelike vacuum with $V=V'=0$, where the prime denotes differentiation with respect to the potential's argument.

The resulting metric functions are
\begin{equation}
f(r)=1-\frac{2M}{r}, \qquad h(r)=\frac{1+\mathfrak l}{1-2M/r}, \qquad \mathfrak l=\xi b^2>-1.
\end{equation}
The horizon remains at $\rh=2M$, while
\begin{equation}
c_h=\sqrt{1+\mathfrak l}, \qquad \kappa=\frac{1}{4M\sqrt{1+\mathfrak l}}.
\end{equation}
We place the independent Maxwell field $A_\mu$ and charged matter of the preceding sections on this fixed metric. They couple minimally to the metric and have no direct coupling to $B_\mu$. Their action therefore retains the internal $U(1)$ symmetry
\begin{equation}
\delta_\eps A_\mu=\partial_\mu\eps, \qquad \delta_\eps\Phi=\ii q\eps\Phi, \qquad \delta_\eps B_\mu=0.
\end{equation}
This symmetry belongs to the Maxwell probe and does not require a corresponding gauge symmetry of the bumblebee field.

To determine whether the same large gauge transformations survive at the horizon, we impose the radial gauge and falloffs used above. Preserving $A_r=0$ and $A_u=\mathcal O(r-\rh)$ requires $\partial_r\eps=\partial_u\eps=0$, leaving an arbitrary smooth function $\eps(z,\bar z)$. The preferred vector imposes no further restriction on this parameter. With the electric normalization $A_u^0=(\rh^2/c_h)\partial_rA_u|_{\cH}$, the explicit dependence on $c_h$ also cancels from the horizon constraint. Equation~\eqref{eq-constraint} and the corresponding surface-charge construction therefore retain their previous forms. The endpoint conditions remain compatible with these transformations because the endpoint phases are free to vary.

The deformation enters the isotropic representation through the relation between the radial coordinate and the lapse. For direct comparison with Schwarzschild, choose $\rho_H=M/2$. Integrating the isotropic coordinate relation gives
\begin{equation}
\rho=\frac{M}{2} \left[ \frac{r-M+\sqrt{r(r-2M)}}{M} \right]^{\sqrt{1+\mathfrak l}}.
\end{equation}
For $R=\rho-\rho_H>0$, this implies
\begin{equation}
A(R)=\frac{R}{M\sqrt{1+\mathfrak l}}+\mathcal O(R^2), \qquad B_H=\frac{\rh}{\rho_H}=4.
\end{equation}
The conversion coefficient is consequently
\begin{equation}
\frac{\mathcal K(R)}{4\ii\omega}=\frac{R}{512\ii\omega M^2\sqrt{1+\mathfrak l}}+\mathcal O(R^2/\omega),\qquad\eta=+1.
\end{equation}
The coordinate transformation and these coefficients recover the Schwarzschild results when $\mathfrak l=0$. With the same endpoint, auxiliary-charge and crossing conditions as above, Eq.~\eqref{eq-local-theorem} gives the single-emission amplitude by multiplying this coefficient by $g^2S(w)\vev{\mathrm{out}|\cS|\mathrm{in}}$. Explicitly, the angular single-emission formula is
\begin{equation}
\vev{\mathrm{out}|a_+(\omega,w;R)\cS|\mathrm{in}} \simeq\frac{g^2R}{512\ii M^2\sqrt{1+\mathfrak l}\,\omega} S(w)\vev{\mathrm{out}|\cS|\mathrm{in}}.
\label{eq-bumblebee-emission}
\end{equation}
For the momentum representation
\begin{equation}
\frac{\mathcal K(R)}{A(R)}=C+\mathcal O(R),\quad \text{with}\quad C=\frac1{128M}.
\end{equation}
The leading prefactor is $g^2/(512\ii\omega M)$. The explicit factor $\sqrt{1+\mathfrak l}$ cancels between $\mathcal K(R)$ and $A(R)$, while the deformation remains in the local momentum contractions through the metric functions. The unchanged prefactor therefore does not remove the background dependence of the full soft factor.

The deformation thus changes the surface gravity and the isotropic conversion, while the normalized horizon soft relation retains the same angular factor and coupling as in Schwarzschild. This example shows that spontaneous Lorentz symmetry breaking in the gravitational background does not by itself obstruct the large gauge symmetry or the leading horizon soft theorem of an independent, minimally coupled Maxwell field.

\paragraph{A background with nonconstant $c(r)$.}

The bumblebee solution considered above still has constant $c$. To test whether this property is essential, we consider a static, spherically symmetric black hole surrounded by matter. A convenient example is a special case of the ``dirty black hole'' geometry \cite{Visser:1992qh,Medved:2003rg}, which can be viewed as a solution of Einstein gravity coupled to an anisotropic matter source. We take
\begin{equation}
f(r)=e^{2\psi(r)}\left(1-\frac{2M}{r}\right), \qquad h(r)=\left(1-\frac{2M}{r}\right)^{-1}, \qquad c(r)=e^{\psi(r)},
\end{equation}
where $\psi(r)$ is smooth and finite at the horizon $r=2M$. For a generic nonconstant $\psi(r)$, 
the geometry can be interpreted as a black hole surrounded by anisotropic matter. The familiar cases can be recovered when $\psi$ is constant, up to a constant rescaling of the time coordinate.

The horizon remains at $\rh=2M$. Since the spatial metric is unchanged, the Schwarzschild isotropic coordinate applies, with $\rho_H=M/2$. Writing $R=\rho-\rho_H>0$, we obtain
\begin{equation}
A(R)=e^{\psi(r(R))}\frac{R}{R+M}, \qquad B(R)=\frac{4(R+M)^2}{(2R+M)^2}.
\end{equation}
With $\psi_h=\psi(2M)$, the horizon data and conversion coefficient are
\begin{equation}
\begin{aligned}
c_h&=e^{\psi_h},\qquad\kappa=\frac{e^{\psi_h}}{4M},\\
\frac{\mathcal K(R)}{4\ii\omega}&=\frac{e^{\psi_h}R}{512\ii\omega M^2}+\mathcal O(R^2/\omega).
\end{aligned}
\end{equation}
and $\eta=+1$. Only $\psi_h$ enters the leading coefficients; radial derivatives of $\psi$ contribute at higher orders. Under the same endpoint, auxiliary-charge and crossing conditions, the single-emission amplitude again follows from Eq.~\eqref{eq-local-theorem}. The displayed coefficient multiplies $g^2S(w)$ and the hard amplitude. The angular single-emission formula is
\begin{equation}
\vev{\mathrm{out}|a_+(\omega,w;R)\cS|\mathrm{in}} \simeq\frac{g^2e^{\psi_h}R}{512\ii M^2\omega} S(w)\vev{\mathrm{out}|\cS|\mathrm{in}}.
\label{eq-redshift-emission}
\end{equation}
The momentum conversion satisfies
\begin{equation}
\frac{\mathcal K(R)}{A(R)}=C+\mathcal O(R),\qquad C=\frac1{128M}.
\end{equation}
The leading momentum prefactor is $g^2/(512\ii\omega M)$. The explicit redshift factor cancels from this prefactor, while its dependence remains in the local momentum contractions through $A(R)$.

Setting $\psi(r)=0$ recovers the Schwarzschild results. A nonconstant redshift deformation changes the isotropic conversion but preserves the normalized horizon soft relation, explicitly demonstrating that constant $f(r)h(r)$ is unnecessary.

\subsection{Yang-Mills extension}
\label{sec-yang-mills}

For a non-Abelian gauge field, the hard flux includes the color carried by gauge radiation itself. The horizon Ward identity must therefore account for both matter and gluons. We illustrate this extension with a Yang-Mills probe coupled to a massless complex scalar, working at tree level about the trivial gauge background.

The canonically normalized action is
\begin{equation}
S_{\mathrm{YM}} =-\int\dd^4x\sqrt{-g} \left[\frac14F_{\mu\nu}^aF^{a\mu\nu} +(D_\mu\Phi)^\dagger D^\mu\Phi\right],
\end{equation}
where
\begin{equation}
\begin{aligned}
F_{\mu\nu}^a &=\partial_\mu A_\nu^a-\partial_\nu A_\mu^a +g_{\mathrm{YM}}f^{abc}A_\mu^bA_\nu^c,\\
D_\mu &=\nabla_\mu-\ii g_{\mathrm{YM}}A_\mu^aT^a, \qquad [T^a,T^b]=\ii f^{abc}T^c.
\end{aligned}
\end{equation}
We take the scalar in the adjoint representation, with $(T^a)_{bc}=-\ii f^{abc}$. The matter current is normalized so that the gauge-field equation reads $(D_\mu F^{\mu\nu})^a=g_{\mathrm{YM}}J^{a\nu}$.

With $x=r-\rh$, we impose
\begin{equation}
A_r^a=0, \qquad A_u^a=\mathcal O(x), \qquad A_{z,\bar z}^a=\mathcal O(1), \qquad \Phi=\mathcal O(1).
\end{equation}
The residual transformations are $\delta_\eps A_\mu^a=(D_\mu\eps)^a$, with $\eps^a=\eps^a(z,\bar z)$. As in the Maxwell theory, their surface charge is determined by the horizon electric field. Defining
\begin{equation}
A_u^{a0}=\frac{\rh^2}{c_h}\left.\partial_rA_u^a\right|_{r=\rh}, \qquad J_u^{a0}=\rh^2\left.J_u^a\right|_{r=\rh},
\end{equation}
the charge on a horizon sphere, in the retarded reference orientation, is
\begin{equation}
\mathcal Q_\eps(u) =\int\dd^2z\,\gamma_{z\bar z}\eps^a A_u^{a0}.
\end{equation}

The radial component of the field equation controls the evolution of this charge. At the horizon it gives
\begin{equation}
\begin{aligned}
\partial_uA_u^{a0} ={}&-\gamma_{z\bar z}^{-1}\partial_u \left(\partial_zA_{\bar z}^{a0}+\partial_{\bar z}A_z^{a0}\right) -g_{\mathrm{YM}}J_u^{a0}\\
&-g_{\mathrm{YM}}\gamma_{z\bar z}^{-1}f^{abc} \left(A_z^{b0}\partial_uA_{\bar z}^{c0} +A_{\bar z}^{b0}\partial_uA_z^{c0}\right).
\end{aligned}
\end{equation}
The explicit metric factors cancel in precisely the same way as in Eq.~\eqref{eq-constraint}. Integrating along the branch yields $Q_\eps=Q_{\mathrm{soft}}+Q_{\mathrm{hard}}+Q_{{\rm end},\eps}$, with
\begin{equation}
\begin{aligned}
Q_{\mathrm{soft}} ={}&\int_{\cH}\dd u\dd^2z\,\eps^a\partial_u \left(\partial_zA_{\bar z}^{a0}+\partial_{\bar z}A_z^{a0}\right),\\
Q_{\mathrm{hard}} ={}&g_{\mathrm{YM}}\int_{\cH}\dd u\dd^2z\,\eps^a \Bigl[\gamma_{z\bar z}J_u^{a0}\\
&\hspace{2.5cm} +f^{abc}\left(A_z^{b0}\partial_uA_{\bar z}^{c0} +A_{\bar z}^{b0}\partial_uA_z^{c0}\right)\Bigr].
\end{aligned}
\end{equation}
The soft charge is linear in the gauge field. The hard charge contains both the matter current and the gluon color flux, the latter vanishing in the Abelian theory. This additional term is what allows the charge to act on hard gluons as well as on matter.

For the soft theorem, we set the matched electric data contributing to $Q_{{\rm end},\eps}^{\pm}$ to zero. This choice is preserved by residual color rotations. The endpoint connections remain pure gauge and free to transform. Their nonlinear flatness condition must be retained when reducing the soft charge.
\begin{equation}
\partial_z A_{\bar z}^{a0}-\partial_{\bar z}A_z^{a0} +g_{\mathrm{YM}}f^{abc}A_z^{b0}A_{\bar z}^{c0}=0\quad\text{at the endpoints}.
\end{equation}
Writing $\Delta X=X(+\infty)-X(-\infty)$ on the reference branch, the pole kernel gives
\begin{equation}
\begin{aligned}
Q_{\mathrm{soft}}^a(w)&=4\pi\Delta A_w^{a0}+Q_{\mathrm{corner}}^a(w),\\
Q_{\mathrm{corner}}^a(w)&=-g_{\mathrm{YM}}f^{abc}\int\dd^2z\, \frac{\Delta(A_z^{b0}A_{\bar z}^{c0})}{w-z}.
\end{aligned}
\label{eq-ym-corner}
\end{equation}
The first term defines $Q_{\mathrm{linear}}^a$. The quadratic endpoint term is distinct from the electric endpoint charge and need not vanish at tree level. We retain it in the charge balance and restrict the formulas below to matrix elements satisfying
\begin{equation}
\langle\mathrm{out}|Q_{\mathrm{corner}}^{a,+}\cS-\cS Q_{\mathrm{corner}}^{a,-}|\mathrm{in}\rangle=0.
\label{eq-ym-corner-sector}
\end{equation}
This is an additional endpoint-sector assumption. Without it, the same matrix element remains on the right-hand side of the soft Ward identity with a minus sign, like the contribution from other boundaries.

The quadratic Yang-Mills action has the same symplectic normalization as a Maxwell field with $g=1$. Its horizon mode expansion is therefore Eq.~\eqref{eq-horizon-mode} with an additional color index. Selecting one color component and using the same distributional kernel $\eps(z,\bar z)=(w-z)^{-1}$, the linear part of the soft charge in the retarded reference orientation becomes
\begin{equation}
Q_{\mathrm{linear}}^a(w) =-\frac1{\sqrt2(1+w\bar w)} \lim_{\omega\to0^+}\omega \left[b_+^a(\omega,w)+b_-^{a\dagger}(\omega,w)\right].
\end{equation}
The hard charge acts on an incoming state as
\begin{equation}
Q_{\mathrm{hard}}^-|\mathrm{in}\rangle =g_{\mathrm{YM}}\sum_{k\in\mathrm{in}} \eps^a(w_k,\bar w_k)T_k^a|\mathrm{in}\rangle,
\end{equation}
with the analogous action on the outgoing state. Here $T_k^a$ acts on the color index of the corresponding particle, including the adjoint color index of a hard gluon.

These two charge actions determine the soft relation. For the amplitude with its external color indices retained, define the action of $S^a(w)$ by
\begin{equation}
\begin{aligned}
&S^a(w)\vev{\mathrm{out}|\cS|\mathrm{in}}\\
&\quad=\frac{1+w\bar w}{\sqrt2\omega}\Biggl[\sum_{l\in\mathrm{out}} \frac{\vev{\mathrm{out}|T_l^a\cS|\mathrm{in}}}{w-w_l} -\sum_{k\in\mathrm{in}} \frac{\vev{\mathrm{out}|\cS T_k^a|\mathrm{in}}}{w-w_k} \Biggr].
\end{aligned}
\end{equation}
Thus the angular kernel is unchanged, while the electric charges are replaced by generators acting on the amplitude. Here $S^a(w)$ acts linearly on the external color indices; it is not a scalar factor obtained by dividing two matrix elements. Under Eq.~\eqref{eq-ym-corner-sector}, the soft vacuum conditions and the sign convention used for the photon, the Ward identity with $\Delta_\eps=0$ gives
\begin{equation}
\bra{\mathrm{out}} b_+^a(\omega,w)\cS-\cS b_-^{a\dagger}(\omega,w) \ket{\mathrm{in}} \simeq 2\eta g_{\mathrm{YM}}S^a(w)\vev{\mathrm{out}|\cS|\mathrm{in}}.
\end{equation}
This is the color counterpart of Eq.~\eqref{eq-soft-theorem}. It has the same large-gauge Ward structure as the familiar soft gluon theorem and its Schwarzschild and dS horizon realizations \cite{Strominger:2013lka,Cheng:2022xgm,Mao:2023rca}. If the soft crossing relation holds, the single-emission formula is
\begin{equation}
\vev{\mathrm{out}|b_+^a(\omega,w)\cS|\mathrm{in}}\simeq \eta g_{\mathrm{YM}}S^a(w)\vev{\mathrm{out}|\cS|\mathrm{in}}.
\end{equation}

Finally, the isotropic representation uses the same normalization convention and geometric conversion as the photon. For the canonical Yang-Mills field, the mode dictionary reads
\begin{equation}
b_\alpha^a(\omega,w;R) =\frac{4\ii\omega}{\mathcal K(R)}a_\alpha^a(\omega,w;R).
\end{equation}
Taking the soft limit before removing the stretched horizon gives
\begin{equation}
\vev{\mathrm{out}|a_+^a(\omega,w;R)\cS|\mathrm{in}} \simeq\eta g_{\mathrm{YM}}\frac{\mathcal K(R)}{4\ii\omega} S^a(w)\vev{\mathrm{out}|\cS|\mathrm{in}}.
\label{eq-ym-local}
\end{equation}

The momentum form follows from Eq.~\eqref{eq-momentum-map} with electric charges replaced by color generators. For the same local state matching and endpoint conditions
\begin{equation}
\begin{aligned}
&\vev{\mathrm{out}|a_+^a(\omega,w;R)\cS|\mathrm{in}}\\
&\quad\simeq\frac{\eta g_{\mathrm{YM}}\mathcal K(R)}{4\ii\omega A(R)} \Biggl[\sum_{l\in\mathrm{out}}\frac{k_{l\mu}\varepsilon^{+\mu}}{k_{l\nu}p^\nu} \vev{\mathrm{out}|T_l^a\cS|\mathrm{in}}\\
&\hspace{3cm}-\sum_{k\in\mathrm{in}}\frac{k_{k\mu}\varepsilon^{+\mu}}{k_{k\nu}p^\nu} \vev{\mathrm{out}|\cS T_k^a|\mathrm{in}}\Biggr].
\end{aligned}
\label{eq-ym-momentum}
\end{equation}
The generators act in the representation of each external particle, with the adjoint representation for hard gluons. The leading momentum prefactor is $\eta g_{\mathrm{YM}}C/(4\ii\omega)$, so the same geometric coefficients apply. Hence the geometric coefficients obtained for the Maxwell probe apply unchanged to Yang-Mills theory, while the hard charge includes the color carried by both matter and gluons.

\paragraph{Summary of the results}

Table~\ref{tab-background-coefficients} lists the leading geometric coefficients in the angular and local momentum representations. The isotropic horizon radius $\rho_H$ specifies the coordinate normalization, and $\eta$ records the boundary orientation. The last two columns give the leading values of $\mathcal K(R)/R$ and $C=\rho_H^3/\rh^4$, respectively. The former determines the radial dependence in Eq.~\eqref{eq-local-theorem}; the latter is the horizon limit of $\mathcal K(R)/A(R)$ in Eq.~\eqref{eq-momentum-theorem-finite-R}. All entries refer to nondegenerate horizons.

For Maxwell single emission, the angular column multiplies $\eta g^2R S(w)/(4\ii\omega)$ and the hard amplitude. The momentum column multiplies $\eta g^2/(4\ii\omega)$, the momentum factor in Eq.~\eqref{eq-momentum-theorem}, and the same hard amplitude. For Yang-Mills theory, $g^2$ is replaced by $g_{\mathrm{YM}}$, and the angular charges and momentum factors act through color generators as in Eqs.~\eqref{eq-ym-local} and \eqref{eq-ym-momentum}. Both representations use the same normalization of the emitted mode. At fixed Killing energies and matched directions, the momentum contraction scales as $A(R)/\omega$, so both forms have leading behavior $R/\omega^2$.

\begin{table*}[!htbp]
\centering
\small
\setlength{\tabcolsep}{3pt}
\renewcommand{\arraystretch}{1.4}
\caption{Geometric coefficients for the horizon soft relations.}
\label{tab-background-coefficients}
\begin{ruledtabular}
\begin{tabular}{@{}>{\raggedright\arraybackslash}p{0.29\textwidth}
>{\centering\arraybackslash}p{0.09\textwidth}
>{\centering\arraybackslash}p{0.05\textwidth}
>{\centering\arraybackslash}p{0.28\textwidth}
>{\centering\arraybackslash}p{0.20\textwidth}@{}}
Background (horizon) & $\rho_H$ & $\eta$ & $\mathcal K(R)/R$ & $C$\\[12pt]
\colrule
Schwarzschild  & $M/2$ & $+1$ & $\displaystyle \frac1{128M^2}$ & $\displaystyle \frac1{128M}$\\[15pt]
S-AdS & $r_S/4$ & $+1$ & $\displaystyle \frac{1+3r_S^2/\ell^2}{32r_S^2}$ & $\displaystyle \frac1{64r_S}$\\[15pt]
dS (cosmological horizon) & $\ell$ & $-1$ & $\displaystyle \frac1{\ell^2}$ & $\displaystyle \frac1\ell$\\[15pt]
RN (outer horizon) & $\sigma/2$ & $+1$ & $\displaystyle \frac{\sigma^3}{4r_+^5}$ & $\displaystyle \frac{\sigma^3}{8r_+^4}$\\[15pt]
S-dS (black hole horizon) & $r_S/4$ & $+1$ & $\displaystyle \frac{1-\Lambda r_S^2}{32r_S^2}$ & $\displaystyle \frac1{64r_S}$\\[15pt]
S-dS (cosmological horizon) & $r_C$ & $-1$ & $\displaystyle \frac{\Lambda r_C^2-1}{2r_C^2}$ & $\displaystyle \frac1{r_C}$\\[15pt]
Bumblebee black hole & $M/2$ & $+1$ & $\displaystyle \frac1{128M^2\sqrt{1+\mathfrak l}}$ & $\displaystyle \frac1{128M}$\\[15pt]
Dirty black hole with $c(r)$ & $M/2$ & $+1$ & $\displaystyle \frac{e^{\psi_h}}{128M^2}$ & $\displaystyle \frac1{128M}$\\[15pt]
\end{tabular}
\end{ruledtabular}
\par\medskip
\end{table*}

The table summarizes how different horizon geometries enter the same momentum-space soft theorem. The coefficient $C$ fixes the geometric prefactor in the chosen isotropic normalization, while $\mathcal K(R)$ relates the projected photon coefficient to the normalized horizon mode. The momentum kernel retains its common charge-weighted form, with the lapse dependence contained in the local momentum contractions and the boundary orientation specified by $\eta$. A constant rescaling of $\rho$ changes $C$ and the projected mode by the same factor, leaving the normalized horizon insertion unchanged. The examples with independent metric functions further demonstrate that the construction does not require constant $f(r)h(r)$. In the bumblebee background, the large gauge symmetry and soft relation also survive gravitational Lorentz symmetry breaking for the independent, minimally coupled Maxwell probe.

With the endpoint electric contributions canceled in the chosen matched sector, the homogeneous Ward identity additionally requires $\Delta_\eps=0$. The remaining boundaries include null infinity for Schwarzschild and RN, the timelike conformal boundary for S-AdS, and the other horizon for S-dS. If a transition has a nonzero auxiliary charge insertion, this contribution must be retained in Eq.~\eqref{eq-ward}. It modifies the soft charge balance without changing the local geometric coefficients listed in the table.

\section{Conclusion and discussion}
\label{sec-discussion}

We have developed a unified formulation of leading soft photon theorems for nondegenerate static spherical horizons. The near-horizon Maxwell constraint takes a common form even when the temporal and radial metric functions are independent and $c(r)=\sqrt{f(r)h(r)}$ varies with radius. Normalizing the photon modes through the horizon symplectic form then yields the same angular soft factor for black hole and cosmological horizons, with their relative sign determined by boundary orientation. The result provides a common description of near-horizon soft emission across this class of backgrounds.

The charge construction specifies the boundary conditions under which this common soft relation holds. Retaining the endpoint electric datum makes the distinction between the complete surface charge and the integrated soft and hard fluxes explicit. In matched fixed electric sectors, the endpoint contributions cancel between the two horizon branches. The homogeneous horizon Ward identity additionally requires $\Delta_\eps=0$, while a nonzero contribution from other boundaries remains part of the global charge balance. The Ward identity directly fixes the combination of outgoing and incoming soft insertions; its reduction to a single-emission formula requires the additional soft crossing relation.

The isotropic representation makes the geometric coefficients in explicit examples accessible without changing the normalized horizon theorem. The angular form is governed by $\mathcal K(R)$, while the local momentum form contains $\mathcal K(R)/A(R)$, with leading value $C=\rho_H^3/\rh^4$. These are geometric factors for the same emitted mode. Their coordinate rescaling is compensated by that of the projected Fourier coefficient. The momentum form requires matching horizon states to local massless momentum modes, including the branch-dependent angular and polarization conventions. The order of limits is fixed by taking the soft limit first at a fixed stretched horizon before the horizon limit. These choices allow the coordinate expressions to be compared consistently while preserving the common angular relation.

The examples test both this normalization and the assumptions underlying the construction. Across the examples, the probe gauge symmetry fixes the common soft structure while the metric determines its geometric coefficients. This separation remains valid for independent metric functions with nonconstant $f(r)h(r)$. Equal momentum prefactors on different backgrounds do not imply equal soft factors, since the local contractions retain their metric dependence. For Yang-Mills theory, under the endpoint conditions including Eq.~\eqref{eq-ym-corner-sector}, the same geometric normalization gives the leading tree-level soft gluon relation about the trivial gauge background. The hard flux now includes the color carried by gluons, and color generators replace electric charges in the angular soft factor.

The integrated horizon constraint also identifies the electromagnetic memory associated with this soft sector. With flat endpoint connections, the zero-frequency radiative mode measures the change in the angular gauge potential. Its divergence is fixed by the integrated matter flux together with the change in the electric datum. The soft insertion and horizon memory thus arise from the same charge-balance law. Connecting this horizon relation to radiation measured elsewhere requires the radial propagation and boundary conditions of the chosen spacetime, including the appropriate greybody factors. Developing that connection would give a concrete setting in which to compare horizon memory with asymptotic observables.

The present analysis provides a starting point for extensions beyond static spherical probe backgrounds. For rotating horizons, the frequency relevant to the soft limit must be defined with respect to the horizon generator, and the angular mode analysis and matching must be adapted accordingly. Extremal horizons require a separate treatment because the nondegenerate expansion used here no longer applies. Extending the charge and normalization analysis to gravitational perturbations, and controlling quantum corrections in the non-Abelian theory, would further test how much of the common horizon soft structure persists beyond the setting studied in this work.

\section*{Acknowledgements}
The authors thank Pujian Mao for earlier collaboration on related topics. This work is supported by the National Natural Science Foundation of China under Grant No.~12405073, the Natural Science Foundation of Tianjin under Grant No.~25JCQNJC01920.

\appendix

\section{Coordinate systems}
\label{sec-coordinates}

We use three coordinate systems adapted to different parts of the calculation. Static coordinates specify the background geometry in terms of the Killing time and areal radius. Null coordinates provide regular charts on the horizon branches, where the characteristic Maxwell constraint determines the charge balance. Isotropic coordinates put the static spatial metric in conformally flat form and are used to compare local Fourier modes in Sec.~\ref{sec-soft-theorem}. These mode conventions are kept separate from the geometric definition of the horizon charge.

\paragraph{Static coordinates.}
We consider a static, spherically symmetric background
\begin{equation}
\dd s^2=-f(r)\dd t^2+h(r)\dd r^2+r^2\dd\Omega_2^2,
\end{equation}
where $t$ parametrizes the static Killing field and $r$ is the areal radius, so each symmetry sphere has area $4\pi r^2$. We use $c(r)=\sqrt{f(r)h(r)}$, where $f$ and $h$ are the independent metric functions in the static chart. Writing $f'(\rh)\ne0$, we use the positive surface gravity associated with $\partial_t$,
\begin{equation}
\kappa=\frac{|f'(\rh)|}{2c_h}.
\end{equation}
Throughout, angular data are expressed in stereographic coordinates
\begin{equation}
\dd\Omega_2^2=2\gamma_{z\bar z}\dd z\dd\bar z, \qquad \gamma_{z\bar z}=\frac{2}{(1+z\bar z)^2}.
\end{equation}
The stereographic coordinates are related to the $(\theta,\phi)$ coordinates by
\begin{equation}
z=e^{\ii\phi}\tan\frac{\theta}{2},\qquad \bar z=e^{-\ii\phi}\tan\frac{\theta}{2}.
\end{equation}

\paragraph{Null coordinates.}
To remove the static coordinate singularity while retaining the areal radius, define
\begin{equation}
r_* = \int^r\sqrt{\frac{h(s)}{f(s)}}\dd s,
\end{equation}
and write the null coordinates as
\begin{equation}
\qquad u=t-r_*,\qquad v=t+r_*.
\end{equation}
Thus, we can get
\begin{equation}
\dd u=\dd t-\frac{c(r)}{f(r)}\dd r\,,\qquad \dd v=\dd t+\frac{c(r)}{f(r)}\dd r\,,
\end{equation}
giving the following metric
\begin{eqnarray}
\dd s^2&=&-f(r)\dd u^2-2c(r)\dd u\dd r+r^2\dd\Omega_2^2,\\
\dd s^2&=&-f(r)\dd v^2+2c(r)\dd v\dd r+r^2\dd\Omega_2^2.
\end{eqnarray}
These charts are regular on the past and future black hole horizons, respectively; their roles are reversed at a cosmological horizon. The mixed coefficients $\mp c(r)$ are retained throughout, with no restriction on $c'(r)$.

For matching at the bifurcation sphere, we introduce
\begin{equation}
U=-e^{-\eta\kappa u},\qquad V=e^{\eta\kappa v},
\end{equation}
where $\eta=\operatorname{sgn}(f'(\rh))$ equals $+1$ at a black hole horizon and $-1$ at a cosmological horizon. Since
\begin{equation}
r_*\sim\eta(2\kappa)^{-1}\log|r-\rh|\,,
\end{equation}
$UV$ vanishes linearly in $|r-\rh|$. The regular extension identifies $U=V=0$ as the bifurcation sphere used in Eq.~\eqref{eq-matching}. The signs of the horizon charges are fixed by boundary orientation, as discussed in the main text.

\paragraph{Isotropic coordinates.}
Starting from the static coordinates \eqref{eq-static} and keeping $t$ unchanged, define the isotropic radius $\rho$ by
\begin{equation}
\frac{\dd\rho}{\rho}=\frac{\sqrt{h(r)}}{r}\dd r.
\end{equation}
The metric becomes
\begin{equation}
\dd s^2=-A^2(\rho)\dd t^2 +B^2(\rho)\bigl(\dd\rho^2+\rho^2\dd\Omega_2^2\bigr),
\end{equation}
where we have
\begin{equation}
A(\rho)=\sqrt{f(r(\rho))},\qquad B(\rho)=\frac{r(\rho)}{\rho}.
\end{equation}
These are the local isotropic coordinates. For the local Fourier representation, we use the four-dimensional coordinates
\begin{equation}
x^\mu=\left( t,\frac{\rho(z+\bar z)}{1+z\bar z}, -\frac{\ii\rho(z-\bar z)}{1+z\bar z}, \frac{\rho(1-z\bar z)}{1+z\bar z} \right).
\end{equation}
In this chart, the metric components are
\begin{equation}
g_{\mu\nu}(x)=\operatorname{diag} \bigl[-A^2(\rho),B^2(\rho),B^2(\rho),B^2(\rho)\bigr],
\end{equation}
and the line element reads
\begin{equation}
\dd s^2=g_{\mu\nu}\dd x^\mu\dd x^\nu =-A^2(\rho)\dd t^2+B^2(\rho)\dd \vec{x}^2.
\end{equation}
If the horizon lies at $\rho=\rho_H$, its positive coordinate distance into the static region is $R=\rho-\rho_H$ for a black hole horizon and $R=\rho_H-\rho$ for a cosmological horizon. We write $A(R)$ and $B(R)$ for evaluation at the corresponding $\rho$. Near a nondegenerate horizon, $r-\rh=\mathcal O(R^2)$, as made explicit in Eq.~\eqref{eq-isotropic}. The freedom to rescale $\rho$ by a positive constant changes the local Fourier conventions but leaves the normalized horizon modes unchanged; the corresponding transformations are given after Eq.~\eqref{eq-isotropic-rescaling}.

\section{Constraint integration and normalization}
\label{app-normalization}

This appendix supplies the intermediate steps behind the charge construction and the soft-mode normalization. We retain the radial integral, the endpoint and zero-mode normalization checks, and the isotropic conversion convention. Throughout, the metric functions remain independent, with $c(r)=\sqrt{f(r)h(r)}$.

\paragraph{Radial constraints and the endpoint charge.}

In radial gauge, the radial Maxwell constraint is
\begin{equation}
\partial_r\left(\frac{r^2}{c}\partial_rA_u\right) =\gamma_{z\bar z}^{-1}\partial_r \left(\partial_zA_{\bar z}+\partial_{\bar z}A_z\right) +g^2r^2j_r.
\end{equation}
Integrating the radial Maxwell constraint with $A_u(\rh)=0$ and $(r^2/c)\partial_rA_u|_{\rh}=A_u^0$ gives
\begin{equation}
\begin{aligned}
A_u(r)=\int_{\rh}^{r}\frac{c(s)}{s^2}\dd s\, \Bigl\{&A_u^0 +\gamma_{z\bar z}^{-1} \left[\partial_z\bigl(A_{\bar z}(s)-A_{\bar z}^0\bigr) +\partial_{\bar z}\bigl(A_z(s)-A_z^0\bigr)\right]\\
&+g^2\int_{\rh}^{s}\dd s'\,s'^2j_r(s')\Bigr\}.
\end{aligned}\label{eq-radial-potential}
\end{equation}
Thus the angular fields, the bulk current and $A_u^0$ determine the radial continuation of $A_u$. For the allowed smooth currents $j_r=\mathcal O(1)$, the current term starts at order $(r-\rh)^2$. The arguments $(u,z,\bar z)$ are held fixed in the radial integrals.

The radial continuation of the matter flux follows separately from current conservation. The equation $\nabla_\mu j^\mu=0$ gives
\begin{equation}
\begin{aligned}
\partial_r(r^2j_u) ={}&-\partial_u(r^2j_r) +\partial_r\left(\frac{r^2f}{c}j_r\right)\\
&+c\gamma_{z\bar z}^{-1} \left(\partial_zj_{\bar z}+\partial_{\bar z}j_z\right).
\end{aligned}
\end{equation}
For any horizon datum $X$, write $\Delta X=X(+\infty)-X(-\infty)$. Integrating Eq.~\eqref{eq-constraint} along the branch gives
\begin{equation}
\Delta A_u^0 =-\gamma_{z\bar z}^{-1} \left(\partial_z\Delta A_{\bar z}^0+\partial_{\bar z}\Delta A_z^0\right) -g^2\int_{-\infty}^{+\infty}\dd u\,j_u^0.
\end{equation}
Multiplying by $\gamma_{z\bar z}\eps/g^2$ and integrating over the sphere recovers the completed charge in the main text. The flux therefore fixes the change in the surface charge, leaving one endpoint electric datum independent. In the fixed electric sector used in the main text, this endpoint contribution is retained in the charge but has vanishing variation.

\paragraph{Charge normalization and the soft insertion.}

The radiative photon bracket in Eq.~\eqref{eq-radiative-bracket} fixes the nonzero-frequency normalization. Indeed, substituting the mode expansion and commutator of Eqs.~\eqref{eq-horizon-mode} and \eqref{eq-mode-bracket}, with $\delta_\Omega^2(z,w)=\gamma_{w\bar w}^{-1}\delta^2(z-w)$, reproduces that bracket using
\begin{equation}
\int_0^\infty\frac{\dd\omega}{\omega}\sin\bigl[\omega(u-u')\bigr] =\frac{\pi}{2}\operatorname{sgn}(u-u').
\end{equation}
The action of the large gauge charge also requires the endpoint modes. With the flat endpoint condition
\begin{equation}
A_A^0(\pm\infty)=\partial_A\phi_\pm, \qquad \Delta A_A^0=\partial_A(\phi_+-\phi_-).
\end{equation}
Following the continuous endpoint prescription of Ref.~\cite{He:2014cra}, the brackets needed for the generator are
\begin{equation}
[\phi_\pm(z,\bar z),A_w^0(u,w,\bar w)] =\mp\frac{\ii g^2}{8\pi}\frac1{z-w},
\end{equation}
together with their complex conjugates. The soft charge is consequently
\begin{equation}
Q_{\mathrm{soft}} =\frac2{g^2}\int\dd^2z\,\eps\, \partial_z\partial_{\bar z}(\phi_+-\phi_-).
\end{equation}
Using the distributional convention $\partial_{\bar z}(z-w)^{-1}=2\pi\delta^2(z-w)$ gives
\begin{equation}
[Q_{\mathrm{soft}},A_w^0] =-\frac{\ii}{2\pi}\int\dd^2z\,\eps\, \partial_z\partial_{\bar z}\frac1{z-w}=\ii\partial_w\eps.
\end{equation}
Thus the endpoint prescription supplies the full gauge transformation of the horizon field.

The matter normalization can be checked independently. For a canonically normalized complex scalar, the matter part of Eq.~\eqref{eq-symplectic} gives
\begin{equation}
[\bar\Phi^0(u,z),\Phi^0(u',w)] =-\frac{\ii}{4\rh^2}\gamma_{w\bar w}^{-1} \operatorname{sgn}(u-u')\delta^2(z-w).
\end{equation}
Since $A_u$ vanishes on the horizon, the current there is
\begin{equation}
j_u^0=\ii q\rh^2 \left(\bar\Phi^0\partial_u\Phi^0-\Phi^0\partial_u\bar\Phi^0\right).
\end{equation}
Integrating by parts for wavepackets that vanish at both endpoints gives $[Q_{\mathrm{hard}},\Phi^0]=-q\eps\Phi^0$, as in Eq.~\eqref{eq-gauge-action}. For $\eps=(w-z)^{-1}$, the flat endpoint condition and $\partial_{\bar z}\eps=-2\pi\delta^2(z-w)$ give Eq.~\eqref{eq-soft-endpoint}. Angular integrals are understood after smooth smearing on the sphere, with no physical boundary at a stereographic chart edge.

Differentiating Eq.~\eqref{eq-horizon-mode} and integrating over $u$ produces $2\pi\delta(\omega)$. The positive-frequency integral retains the endpoint weight $\int_0^\infty\dd\omega\,\delta(\omega)=1/2$. Hence
\begin{equation}
\Delta A_z^0 =-\frac{\sqrt2g}{8\pi(1+z\bar z)} \lim_{\omega\to0^+}\omega \left[b_+(\omega,z)+b_-^\dagger(\omega,z)\right]. \label{eq-zero-mode}
\end{equation}
Substitution into Eq.~\eqref{eq-soft-endpoint} yields Eq.~\eqref{eq-soft-charge}. In the homogeneous, matched endpoint sector used in the main text, the Ward identity gives Eq.~\eqref{eq-soft-theorem}. If the assumed soft crossing relation holds, the coefficient for a single emission is half that of the crossed insertion. With the definition in Eq.~\eqref{eq-angular-factor}, the finite soft residue is proportional to $\omega S(w)$.

The zero-frequency insertions are understood as distributions in Ward matrix elements, using angular and frequency smearing. Their soft limits are assumed to exist in the chosen infrared sector. Finite norm at nonzero frequency fixes the mode normalization but does not by itself establish the existence of this limit.

\paragraph{Local Fourier measure and isotropic conversion.}

The isotropic representation is evaluated at a fixed stretched horizon. Here $A$ and $B$ denote their values at that radius. For covariant spatial momentum components $p_i$, write $P^2=\delta^{ij}p_i p_j$. The mass-shell condition is
\begin{equation}
g^{\mu\nu}p_\mu p_\nu =-\frac{E^2}{A^2}+\frac{P^2}{B^2}=0, \qquad \omega=\frac AB P>0.
\end{equation}
Here the integration variable $E=-p_t$ equals the Killing frequency $\omega$ on the positive-frequency mass shell. Resolving the positive-frequency root gives
\begin{equation}
\int\dd E\,2\pi\theta(E) \delta\left(-\frac{E^2}{A^2}+\frac{P^2}{B^2}\right) =\frac{\pi A^2}{\omega}.
\end{equation}
This is the origin of the factor $A^2$ in the coordinate Fourier convention used here. Changing from $P$ to $\omega$ gives
\begin{equation}
A^2\frac{\dd^3p}{2\omega} =\frac{B^3}{2A}\omega\dd\omega\,\dd\Omega_p.
\end{equation}
Canonical normalization is supplied separately by the horizon symplectic form. This measure fixes a local Fourier convention with the metric held fixed. It does not determine radial propagation or an angular projection onto one photon direction; those require mode functions and boundary data. The coefficient $a_\alpha(\omega,w;R)$ used in the main text is the projected coefficient normalized relative to the horizon mode. For the same angular wavepacket, Eq.~\eqref{eq-mode-map} fixes its normalization. A map from unrestricted local Fourier data to this coefficient additionally requires angular projection and radial mode data. Hermitian conjugation gives
\begin{equation}
b_\alpha^\dagger(\omega,w;R) =-\frac{4\ii\omega}{g\mathcal K(R)} a_\alpha^\dagger(\omega,w;R).
\end{equation}
The stretched-horizon representative is chosen to recover the intrinsic soft insertion through
\begin{equation}
\lim_{\omega\to0^+}\omega b_\alpha(\omega,w) =\frac{4\ii}{g} \lim_{R\to0^+}\frac1{\mathcal K(R)} \lim_{\omega\to0^+}\omega^2a_\alpha(\omega,w;R). \label{eq-ordered-limit}
\end{equation}
The frequency limit is taken first at fixed $R$. The outer limit then removes the stretched horizon. This prescription fixes the combination $\omega^2a_\alpha/\mathcal K$ that appears in Eq.~\eqref{eq-local-theorem}.

Finally, the freedom to rescale the isotropic radius provides a direct normalization check. Under $\rho'=\lambda\rho$ with $\lambda>0$,
\begin{equation}
R'=\lambda R, \qquad B'(R')=\frac{B(R)}{\lambda}, \qquad \mathcal K'(R')=\lambda^3\mathcal K(R), \qquad C'=\lambda^3C.
\label{eq-isotropic-rescaling}
\end{equation}
For the same angular projection, $a_\alpha'(\omega,w;R')=\lambda^3a_\alpha(\omega,w;R)$, so the normalized horizon operator and Eq.~\eqref{eq-ordered-limit} remain unchanged. The conversion coefficients therefore encode the isotropic normalization, while the horizon soft insertion retains the normalization fixed by the symplectic form.

\end{document}